\documentclass[letterpaper,english,aps,prx,superscriptaddress,twocolumn,amsfonts, amssymb]{revtex4}
\usepackage[T1]{fontenc}
\pdfoutput=1
\usepackage{textcomp}
\usepackage{mathrsfs}
\usepackage{amsmath}
\usepackage{amssymb}
\usepackage{graphicx}
\usepackage{esint}
\usepackage{wasysym}
\usepackage[normalem]{ulem}

\makeatletter
\usepackage[english]{babel}

\usepackage[bookmarks=true,colorlinks,linkcolor=blue,urlcolor=blue,citecolor=blue]{hyperref}

\usepackage{epsfig,graphicx,psfrag,amsmath,amssymb,float}
\usepackage[caption=false]{subfig}

\@ifundefined{showcaptionsetup}{}{%
 \PassOptionsToPackage{caption=false}{subfig}}
\usepackage{orcidlink}
\usepackage{subfig}
\makeatother

\usepackage{babel}
\begin{document}
\title{Quantum liquids of the S=3/2 Kitaev honeycomb and related Kugel-Khomskii models}

\author{W. M. H. Natori \orcidlink{0000-0002-0740-2956}}
\affiliation{Institute Laue-Langevin, BP 156, 41 Avenue des Martyrs, 38042 Grenoble
Cedex 9, France}
\affiliation{Blackett Laboratory, Imperial College London, London SW7 2AZ, United
Kingdom}
\author{Hui-Ke Jin \orcidlink{0000-0002-0880-6557}}
\affiliation{Department of Physics, TQM, Technische Universität München, 85748
Garching, Germany}
\author{J. Knolle \orcidlink{0000-0002-0956-2419}}
\affiliation{Department of Physics, TQM, Technische Universität München, 85748
Garching, Germany}
\affiliation{Munich Center for Quantum Science and Technology (MCQST), 80799 Munich,
Germany}
\affiliation{Blackett Laboratory, Imperial College London, London SW7 2AZ, United
Kingdom}

\begin{abstract}
The $S=3/2$ Kitaev honeycomb model (KHM) is unique among the spin-$S$ Kitaev models due to a massive ground state quasi-degeneracy that hampered previous numerical and analytical studies. In a recent work~\cite{jin2022unveiling}, we showed how an SO(6) Majorana parton mean-field theory of the $S=3/2$ isotropic KHM explains the anomalous features of this Kitaev spin
liquid (KSL) in terms of an emergent low-energy Majorana flat band. Away from the isotropic limit, the $S=3/2$ KSL generally displays a quadrupolar order with gapped or gapless Majorana excitations, features that were quantitatively confirmed by DMRG simulations. In this paper, we explore the connection between the $S = 3/2$ KHM with Kugel-Khomskii models and discover new exactly soluble examples for the latter. We perform a symmetry analysis for the variational parton mean-field \emph{Ans{\"a}tze} in the spin and orbital basis for different quantum liquid phases of the $S=3/2$ KHM. Finally, we investigate a proposed time-reversal symmetry breaking spin liquid induced by a {[}111{]} single ion anisotropy and elucidate its topological properties as well as experimental signatures, e.g. an unquantized thermal Hall response. 
\end{abstract}
\maketitle

\section{Introduction}

The celebrated $S=1/2$ Kitaev honeycomb model (KHM)~\citep{Kitaev2006} bridges different research fields, i.e., the theory of integrable models, topological quantum computation, and Mott insulators under strong spin-orbit coupling~\citep{Hermanns2018,Takagi2019,Winter2017,TrebstPhysRep2022}. The KHM's eigenstates display exact spin fractionalization into static $Z_{2}$ \emph{fluxes} and Majorana \emph{matter} fermions, resulting in short-range spin correlations characteristic of quantum spin liquids (QSLs)~\citep{Baskaran2007}. Kitaev's original
interest was to instantiate a simple strongly correlated Hamiltonian hosting non-abelian anyon excitations, therefore providing a toy model for fault-tolerant quantum computation~\citep{Kitaev2006}. This initial motivation explains both the surprise and the excitement about the first proposals of KHM implementations in heavy-ion Mott insulators~\citep{Jackeli2009} that later coined the term \emph{Kitaev materials}~\citep{Hermanns2018,Takagi2019,Winter2017,TrebstPhysRep2022}. 

Kitaev materials generally display long-range ordered ground states stabilized by other symmetry-allowed exchanges~\citep{Chaloupka2013,ChaloupkaPRB2015,Rau2014,Winter2016,Gohlke2017,Janssen_2019,Consoli2020,Janssen2016,Janssen2017,Maksimov2020} and intense research has focused on the search for compounds approaching the Kitaev spin liquid (KSL)~\cite{Winter2017,Hermanns2018,motome2020hunting}. One noteworthy example is $\alpha$-RuCl$_{3}$~\citep{PlumbPRB2014}, which transitions from a zigzag ordered state~\citep{Sears2015} to a magnetically disordered phase under the application of a moderate in-plane magnetic field~\citep{Banerjee2016,Banerjee2017,Banerjee2018,BaekPRL2017,Wulferding2020,Wang2020}. The disordered phase is reminiscent of the chiral spin liquid (CSL) predicted by Kitaev~\citep{Kitaev2006}, a point supported by experiments reporting half-quantization of the thermal Hall coefficient~\citep{Kasahara2018,Yokoi2021}, but which is currently under debate~\citep{lefranccois2022evidence,czajka2023planar,bruin2022robustness}.

A recent alternative route to a KSL in $\alpha$-RuCl$_{3}$ was proposed for heterostructures involving monolayers in contact with graphene~\citep{Biswas2019,Leeb2021}. The proximity effect strains the insulator~\citep{Biswas2019} and can enhance the relative importance of Kitaev interactions~\citep{Winter2016,Winter2017}. Another promising direction involves Kitaev materials with 3$d$ magnetic ions~\cite{Liu2018,Sano2018,Liu2020}. As an example, the cobalt-based Kitaev material Na$_{3}$Co$_{2}$SbO$_{6}$~\citep{Songvilay2020}
was proposed to reach the KSL state by reducing its trigonal crystal field through pressure or strain~\citep{Liu2020}. 
The 3$d$ materials were also essential for conceiving higher-spin Kitaev materials with $S>1/2$~\citep{XuNPJ2018,Stavropoulos2019,Xu2020,Stavropoulos2021}. They provide experimental motivation to revisit what were once purely theoretical questions. The spin-$S$ KHMs retain two characteristics of the famous $S=1/2$ case~\citep{Baskaran2008}: i) there is one conserved operator per plaquette defining a static $Z_{2}$ flux, and ii) one can define a Jordan-Wigner transformation and obtain emergent Majorana fermion excitations for half-integer spin $S$. These two characteristics are sufficient to ensure ultra-short ranged spin correlations entailing a QSL ground state~\citep{Baskaran2008}. Nevertheless, these results did not yield an exact solution or a quantitative theory for the Kitaev spin liquids with $S>1/2$. 

An alternative approach is to start from the semiclassical large-$S$ limit~\cite{IoannisNatComm2018}, where the KHM can be mapped onto a toric-code model~\citep{Kitaev2003} over dimers forming a fixed kekule pattern, which provides an adequate understanding of the model for $S>3/2$. The breakdown of this approximation for $S=1/2$ and $S=1$ is interpreted as the formation of QSLs with mobile fractionalized excitations, as evinced by independent numerical studies~\citep{Dong2020,LeePRR2020}. The specific case $S=3/2$ marks the borderline of the stability of the large-$S$ KSL~\citep{IoannisNatComm2018} and has proven to be a challenging numerical problem due to a pile-up of low-energy excitations~\citep{jin2022unveiling}. 

The proposal that $S=3/2$ Kitaev exchanges are relevant for 2D van der Waals magnets~\citep{XuNPJ2018,LebingChen2020,Xu2020,InheeLee2020,Stavropoulos2021} provides a strong experimental motivation to readdress the nature of this exotic QSL. Recently, we tackled this problem by studying the $S=3/2$ KHM in terms of SO(6) Majorana partons~\citep{Wang2009,Corboz2012,Natori2016,Natori2017}.
It allows an \emph{exact} mapping of the $Z_{2}$ fluxes \citep{Baskaran2008} into static $Z_{2}$ gauge operators in analogy to the $S=1/2$ KHM~\citep{jin2022unveiling}. However, despite the presence of a static gauge field the ensuing Majorana problem is fully interacting which prevents a full exact solution. 

\begin{figure}[!t]
    \centering
    \includegraphics[width=0.8\linewidth]{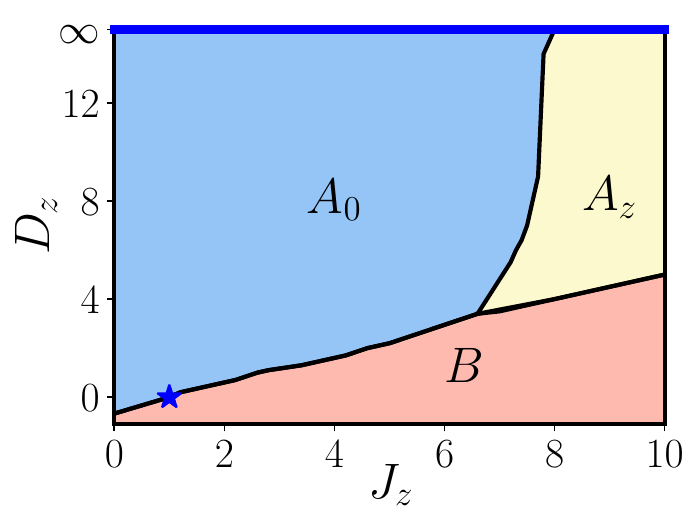}
    \caption{The mean-field ground-state phase diagram of the S=3/2 KHM with a [001] SIA in the zero-ﬂux sector. The $A_0$ phase is a Dirac QSL with spin quadrupolar order $\left\langle T^{z} \right\rangle = Q^z<0$. In the $A_z$ ($B$) phase, the spinon excitations are gapped with $Q^z<0$ ($Q^z>0$). At the isotropic point (blue star), the ground state is a Dirac QSL with $Q^z=0$. The bold blue line at $D_z =\infty$ with $J_z < 8$ ($J_z > 8$) represents the effective gapless (gapped) S=1/2 KSL. The gapless phases in S=3/2 and S=1/2 KHMs can continuously connect to each other through the $A_0$ phase.}
    \label{fig:phasediagramNC}
\end{figure}

A parton mean-field theory (PMFT) of this model perturbed by a flux-conserving {[}001{]} single-ion anisotropy (SIA) unveiled a rich phase diagram with four types of QSLs (see Fig.~\ref{fig:phasediagramNC}): (i) a quantum spin-\emph{orbital} liquid at the isotropic point ($J_{\gamma}=1$), (ii) a gapless QSL dubbed $A_0$ phase adiabatically connected with the $S=1/2$ KSL, (iii) the same as (ii) for the gapped $S=1/2$ KSL, and (iv) a gapped QSL dubbed $B$ phase with vanishingly small flux excitations. The predictions of PMFT are in remarkable and even \emph{quantitative} agreement with state-of-the-art DMRG simulations on $3\times4$ tori and $4\times8$ cylinders. The abundance of low-energy excitations, which hampered previous DMRG simulations of the isotropic KHM, can be attributed to an almost zero-energy flat band of Majorana fermion excitations within the framework of PMFT. 

Our previous work~\citep{jin2022unveiling} also included a perturbative study of the isotropic $S=3/2$ KHM under the {[}111{]} SIA that naturally arises in minimal models of van der Waals magnets~\citep{XuNPJ2018,Xu2020,Stavropoulos2021}. Within the zero-flux sector, this perturbation induces a three-site interaction that in turn leads to a  \emph{spontaneously} time-reversal symmetry (TRS) breaking QSL. This $S=3/2$ KSL thus shares similarities with the celebrated $S=1/2$ chiral KSL induced by a magnetic field~\citep{Kitaev2006} but is distinguished from it by its coexistence with an octupolar order parameter and a zero total Chern number~\citep{jin2022unveiling}.

In this paper, we explore the connection of the $S=3/2$ KHM with Kugel-Khomskii (KK) models by studying the $S=3/2$ operators in terms of pseudo-dipole $\sigma_{i}^{\gamma}$ and pseudo-orbital operators $T_{i}^{\gamma}$~\citep{Natori2016,Romhanyi2017,Natori2017,Natori2018,Farias2020,Yamada2018,Yamada2021}. This facilitates the identification of similarities with integrable KK models~\citep{Yao2009,Nussinov2009,Wu2009,Yao2011,Chua2011,Farias2020,Natori2020,Chulliparambil2020,Seifert2020,Ray2021,Chulliparambil2021,Zhuang2021,WangHaoranPRB2021} and in doing so we discover new soluble examples.
Moreover, the connection to KK models allows for a reinterpretation of the quantum liquid phases. We also provide a symmetry classification of the PMFT and discuss properties of the quantum liquid phases, in particular the one breaking TRS in the presence of the experimentally relevant {[}111{]} SIA.

The paper is structured as follows. Section~\ref{sec:Conserved-Quantities} reviews essential results on the theory of integrable KK models and the spin-$S$ KHM. It then translates these results to the $S=3/2$ case using the pseudo-dipole and pseudo-orbital operators. Section~\ref{sec:Parton-MFT} presents details for the parton representation of an exactly solvable model directly related to the $S=3/2$ KHM. This section also discusses the effects of symmetry constraints on the allowed order parameters and their relations to the properties of the previously uncovered QSLs phases. Section~\ref{sec:CSL} discusses the origins of the first-order phase transition to the TRS breaking $S=3/2$ KSL, as well as its observed topological properties. We conclude in section \ref{sec:Conclusions} with open questions for future research.

\section{Review of some exact results \label{sec:Conserved-Quantities}}

\subsection{Soluble vector models and spin-$S$ KHMs}

We start by recalling a class of exactly solvable spin-$S$ models directly related to the KHM. Consider a set of operators $\Gamma^{a}$ ($a=1,2,...,2q+3,q\in\mathbb{N}_{0}$) defined over a $2^{q+1}$ dimensional Hilbert space which forms a basis for the Clifford algebra
\begin{align}
\left\{ \Gamma_{i}^{a},\Gamma_{i}^{b}\right\}  & =2\delta_{ab},\nonumber \\
\left[\Gamma_{i}^{a},\Gamma_{j}^{b}\right] & =0,\text{ if }i\neq j,\label{eq:Clifford_algebra}
\end{align}
with $i$ and $j$ labeling points on a graph. Several algorithms have been developed to generate models whose Hilbert space is restricted to a sub-algebra whose dimension scales polynomially with the number of lattice bonds~\cite{Nussinov2009,miao2020exact}. In particular, they proposed
the class of \emph{vector models}~\citep{Nussinov2009}
\begin{equation}
H_{\text{vec}}=\sum_{\left\langle ij\right\rangle _{a}}J_{a}\Gamma_{i}^{a}\Gamma_{j}^{a},\label{eq:vector_models}
\end{equation}
in which the label $a$ is assigned at most once for each type of bond in the lattice. All vector models commute with an extensive number of local operators given by an ordered product $\Gamma$ on the elementary plaquettes~\citep{Nussinov2009}. 

\begin{figure}
\begin{centering}
\includegraphics[width=0.4\columnwidth]{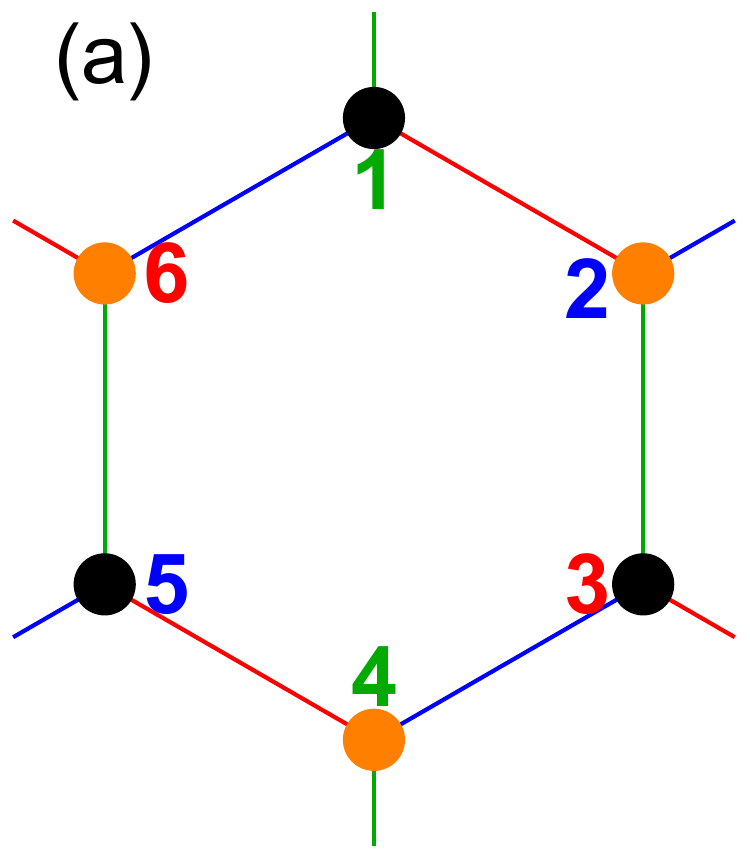}\quad\includegraphics[width=0.4\columnwidth]{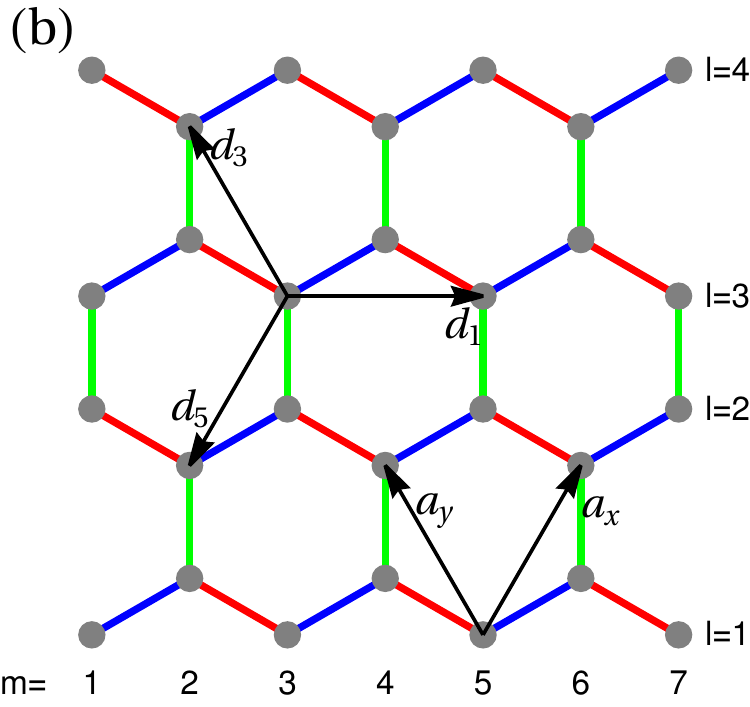}
\par\end{centering}
\caption{\label{fig:plaquette} Conventions for the honeycomb lattice that are used throughout the text. (a) Detail of the honeycomb plaquette. The colors green, blue, and red correspond to $\gamma=z,x,y$, respectively. At each bond $\left\langle ij\right\rangle _{\gamma}$, the interaction
between the spins is given by $J_{\gamma}S_{i}^{\gamma}S_{j}^{\gamma}$,
in which $\gamma$ is defined by the bond. (b) Site counting convention used in the Jordan-Wigner transformation discussed in the text together with the labelling convention of the nearest-neighbor vectors $\mathbf{a}_{x,y}$
and next-nearest-neighbor vectors $\mathbf{d}_{1,3,5}$.}
\end{figure}

An even larger number of integrable models can be defined with the operators $\Gamma^{ab}=\frac{1}{2i}\left[\Gamma^{a},\Gamma^{b}\right]$
($a<b$ and $q\ge1$) \citep{Chulliparambil2020,Chulliparambil2021}.
For concreteness, we express these generalizations only on the honeycomb lattice, where they read
\begin{align}
H & =\sum_{\left\langle ij\right\rangle _{\gamma}}K_{\gamma}\Gamma_{i}^{\gamma}\Gamma_{j}^{\gamma}\nonumber \\
 & +\sum_{\left\langle ij\right\rangle _{\gamma}}\sum_{\alpha=4}^{2q+1}\left(K_{\gamma}^{\alpha}\Gamma_{i}^{\gamma}\Gamma_{j}^{\gamma\alpha}+K_{\gamma}^{\prime\alpha}\Gamma_{i}^{\gamma\alpha}\Gamma_{j}^{\gamma}\right)\nonumber \\
 & +\sum_{\left\langle ij\right\rangle _{\gamma}}\sum_{\alpha,\beta=4}^{2q+1}J_{\gamma}^{\alpha\beta}\Gamma_{i}^{\gamma\alpha}\Gamma_{j}^{\gamma\beta},\label{eq:general_flavor_honeycomb}
\end{align}
with the three bond directions $\gamma$ expressed by different colors in Fig.~\ref{fig:plaquette}. 

Next, we can discuss the connection with the spin-$S$ KHM on the honeycomb lattice given by the Hamiltonian 
\begin{equation}
H_{\text{Kit}}=\sum_{\langle ij\rangle_{\gamma}}J_{\gamma}S_{i}^{\gamma}S_{j}^{\gamma},\label{eq:H_Kitaev}
\end{equation}
in which $\gamma$ labels both the inequivalent bonds on the honeycomb lattice and the corresponding spin quantization axis in the cubic frame~\citep{Janssen_2019,Consoli2020,Maksimov2020}.

The operators $\sigma_{i}^{\gamma}=2S_{i}^{\gamma}$
satisfy the Clifford algebra in Eq.~\eqref{eq:Clifford_algebra} only for $S=1/2$, which thus corresponds to the $q=0$ vector model. The conserved operators for $S=1/2$ are $W_{p}^{1/2}=\sigma_{1}^{z}\sigma_{2}^{x}\sigma_{3}^{y}\sigma_{4}^{z}\sigma_{5}^{x}\sigma_{6}^{y}$~\citep{Kitaev2006} with the label convention set in Fig.~\ref{fig:plaquette}(a). Kitaev then provided an exact solution of the $S=1/2$ model using a Majorana fermion representation 
\begin{equation}
\sigma_{i}^{\gamma}=-i\eta_{i}^{\gamma}\theta_{i}^{0},\label{eq:theta_0_parton}
\end{equation}
in which the four Majorana flavors satisfy $\left\{ \Upsilon_{i}^{\alpha},\Upsilon_{j}^{\beta}\right\} =2\delta_{ij}\delta^{\alpha\beta}$, where $\Upsilon$ is an $\eta$ or $\theta^{0}$ flavor. The Hamiltonian in terms of Majoranas is
\begin{equation}
H_{\text{Kit}}^{S=1/2}=\sum_{\langle ij\rangle_{\gamma}}\frac{J_{\gamma}}{4}\hat{u}_{\langle ij\rangle_{\gamma}}i\theta_{i}^{0}\theta_{j}^{0},\label{eq:S1_2_Majorana}
\end{equation}
in which $\hat{u}_{\langle ij\rangle_{\gamma}}=-i\eta_{i}^{\gamma}\eta_{j}^{\gamma}$
are conserved $Z_{2}$ bond operators akin to a static gauge field. The product of eigenvalues of $\hat{u}_{\langle ij\rangle_{\gamma}}$ around a plaquette fixes the
$\left\{ W_{p}^{1/2}\right\} $ flux sector~\citep{Kitaev2006}. The ground state in the thermodynamic limit is characterized by $W_{p}^{1/2}=+1,\forall p$ \citep{Lieb1994} with a dispersion of the matter sector 
given by
\begin{equation}
\epsilon (\mathbf{k})=\frac{1}{2}\left|J_{z}+J_{x}e^{i\mathbf{k}\cdot\mathbf{a}_{x}}+J_{y}e^{i\mathbf{k}\cdot\mathbf{a}_{y}}\right|,\label{eq:mu_k}
\end{equation}
in which $\mathbf{a}_{x,y}=\pm\frac{1}{2}\hat{\mathbf{x}}+\frac{\sqrt{3}}{2}\hat{\mathbf{y}}$ as shown in Fig.~\ref{fig:plaquette}.

The KHM for $S>1/2$ is not within the class of vector models since the anticommutator $\left\{ S_{i}^{a},S_{i}^{b}\right\} $ corresponds to a quadrupolar operator. Nevertheless, using identities 
\begin{align}
\left\{ e^{i\pi S_{i}^{\alpha}},S_{i}^{\beta}\right\}  & =0\text{ if }\alpha\neq\beta,\nonumber \\
\left[e^{i\pi S_{i}^{\alpha}},S_{i}^{\alpha}\right] & =0,\label{eq:exp_anticomm}
\end{align}
it is still possible to find one conserved operator $W_{p}^{S}$
per plaquette given by~\citep{Baskaran2008}
\begin{equation}
W_{p}^{S}=-\exp\left[i\pi\left(S_{1}^{z}+S_{2}^{x}+S_{3}^{y}+S_{4}^{z}+S_{5}^{x}+S_{6}^{y}\right)\right],\label{eq:Wp_Baskaran}
\end{equation}
in which the minus sign was inserted to include $W_{p}^{1/2}$ as a specific case. 
Since spin operators do not commute with $W_{p}^{S}$ for any $S$, one can prove that spin-spin correlations vanish beyond nearest neighbors and there is no long-range magnetic order in any flux eigenstates of spin-$S$ KHMs~\citep{Baskaran2008}.

The exponential operators in Eq.~\eqref{eq:exp_anticomm} can also be used for defining a Jordan-Wigner-like transformation (JWT) leading to an analytical representation of the $Z_{2}$ flux sector of the spin-$S$ KHM~\citep{Baskaran2008}. The JWT starts with the definition of a string operator
\begin{equation}
\mu_{n}=\prod_{m<n}e^{i\pi\left(S_{m}^{z}+S\right)},\label{eq:string_operator}
\end{equation}
in which $m$, $n$ label the sites following an order defined by strings running over the $xy$ bonds~\citep{FengPRL2007,ChenJPhysA2008,Nasu2014,Miao2019}
[see Fig.~\ref{fig:plaquette}(b)]. At the $n$th site, the exchange interactions along the strings are given by $J_{t_{1}}S_{n-1}^{t_{1}}S_{n}^{t_{1}}$
and $J_{t_{2}}S_{n}^{t_{2}}S_{n+1}^{t_{2}}$, where $t_{1},t_{2}=x,y$.
We can then define
\begin{align}
\xi_{n} & \equiv e^{i\pi\left(S_{n}^{t_{1}}+S\right)}\mu_{n},\nonumber \\
\chi_{n} & \equiv e^{i\pi\left(S_{n}^{t_{2}}+S\right)}\mu_{n},\label{eq:XiChiFermion}
\end{align}
which satisfies Majorana fermion (hard-core boson) statistics for
half-integer (integer) values of $S$. For any pair of sites $ij$ forming a $z$-bond, $u_{ij}=e^{i\pi S}\chi_{i}\chi_{j}$
is a Hermitian operator commuting with the Hamiltonian~\citep{Baskaran2008}, and is directly related to the bond operators $\hat{u}_{\left\langle ij\right\rangle _{z}}$
discussed above, i.e., they can also be used to fix the KHM flux sectors. On the other hand, $\xi_{n}$ represents Majorana fermions for the matter sector only when $S=1/2$, and we need to get into the specifics for understanding KHM with $S>1/2$. 

\subsection{Spin-orbital representation of the Spin-3/2 KHM}

For the remainder, we focus on the $S=3/2$ case and derive an alternative representation in terms of a KK model. We start by defining the spin-3/2 pseudo-dipoles $\boldsymbol{\sigma}$ and pseudo-orbitals $\mathbf{T}$ as follows 
\begin{align}
\sigma_{i}^{\alpha} & =-i\exp\left(i\pi S_{i}^{\alpha}\right),\nonumber \\
T_{i}^{z} & =\left(S_{i}^{z}\right)^{2}-\frac{5}{4},\nonumber \\
T_{i}^{x} & =\frac{1}{\sqrt{3}}\left[\left(S_{i}^{x}\right)^{2}-\left(S_{i}^{y}\right)^{2}\right],\nonumber \\
T_{i}^{y} & =\frac{2\sqrt{3}}{9}\overline{S_{i}^{x}S_{i}^{y}S_{i}^{z}},\label{eq:orbitals_as_spins}
\end{align}
in which the bar indicates a sum over all permutations of the operators under it~\citep{Chen2010}. The definition of $\boldsymbol{\sigma}$ is motivated by the exponential operators in Eqs.\eqref{eq:exp_anticomm},
\eqref{eq:Wp_Baskaran}, and \eqref{eq:string_operator}, and an imaginary factor $-i$ ensures that the pseudo-dipoles satisfy the SU(2) algebra for $S=1/2$ operators. The $T^{z}$ and $T^{x}$ operators are $S=3/2$ quadrupoles that commute with $\boldsymbol{\sigma}$ and transform as $e_{g}$ orbital operators by transformations in real space. Including
the octupolar operator $T^{y}$ which forms a unidimensional representation of the $O_{h}$ group~\citep{Chen2010}, $\mathbf{T}$ also satisfy the SU(2) algebra. The algebra of $\left(\boldsymbol{\sigma},\mathbf{T}\right)$
can be summarized as follows
\begin{align}
\left[\sigma_{i}^{\alpha},\sigma_{j}^{\beta}\right] & =2i\delta_{ij}\epsilon^{\alpha\beta\gamma}\sigma_{i}^{\gamma},\nonumber \\
\left[T_{i}^{\alpha},T_{j}^{\beta}\right] & =2i\delta_{ij}\epsilon^{\alpha\beta\gamma}T_{i}^{\gamma},\nonumber \\
\left\{ \sigma_{i}^{\alpha},\sigma_{j}^{\beta}\right\}  & =\left\{ T_{i}^{\alpha},T_{j}^{\beta}\right\} =2\delta_{ij}\delta^{\alpha\beta},\nonumber \\
\left[\sigma_{i}^{\alpha},T_{j}^{\beta}\right] & =0,\label{eq:algebra_sigma_T}
\end{align}
in which $\epsilon^{\alpha\beta\gamma}$ is the anti-symmetric Levi-Civita
symbol. 
The $\left(\boldsymbol{\sigma},\mathbf{T}\right)$ operators were extensively used in the description of $j=3/2$ Mott insulators as they allow an alternative representation of multipolar interactions and a transparent representation of  global symmetries~\citep{Natori2016,Romhanyi2017,Natori2017,Natori2018,Farias2020,Yamada2018,Yamada2021,Chen2010}.

We can then reformulate the $S=3/2$ KHM after rewriting $S_{i}^{\gamma}$ like~\citep{Natori2017,Farias2020}
\begin{equation}
S_{i}^{\gamma}=-\frac{\sigma_{i}^{\gamma}}{2}-\sigma_{i}^{\gamma}T_{i}^{\alpha\beta},\label{eq:S_gamma}
\end{equation}
in terms of well-known 120$^{\circ}$ compass operators for orbital interactions  $T_{i}^{xy}=T_{i}^{z}$ and $T_{i}^{yz(zx)}=\left(-T_{i}^{z}\pm\sqrt{3}T_{i}^{x}\right)/2$~\citep{Kugel1982,Nussinov2015}. The explicit relationship between the $\left|S^{z}\right\rangle $ and the $\left|\sigma^{z},T^{z}\right\rangle $ basis states is presented in Appendix~\ref{sec:Matrices_sigma_T}. We note that Eq.~\eqref{eq:S_gamma} entails that the $S^{z}=\pm3/2$ ($S^{z}=\pm1/2$) states are the eigenstates of the quadrupolar operator $T^{z}$ with eigenvalue $+1$ ($-1$).

Applying Eq.~\eqref{eq:S_gamma} onto the $S=3/2$ KHM maps it onto a KK model~\citep{Kugel1982,Nussinov2015,Khomskii2020,ChenReview2021}
\begin{equation}
H_{\text{Kit}}=\sum_{\langle ij\rangle_{\gamma}}J_{\gamma}\sigma_{i}^{\gamma}\sigma_{j}^{\gamma}\left(\frac{1}{2}+T_{i}^{\alpha\beta}\right)\left(\frac{1}{2}+T_{j}^{\alpha\beta}\right).\label{eq:Kit_pseudo_dipoles}
\end{equation}

This exact mapping turns out to be very useful for understanding some of the properties of the $S=3/2$ KHM. For example, the commutation $\left[\sigma_{i}^{\alpha},T_{j}^{\beta}\right]=0$ entails
in analogy with the $S=1/2$ KHM that
\begin{equation}
W_{p}^{3/2}=\sigma_{1}^{z}\sigma_{2}^{x}\sigma_{3}^{y}\sigma_{4}^{z}\sigma_{5}^{x}\sigma_{6}^{y}=W_{p}^{\sigma}\label{eq:Wp_sigma}
\end{equation}
commutes with $H_{\text{Kit}}$. The same result is obtained after inserting Eq.~\eqref{eq:orbitals_as_spins} into Eq.~\eqref{eq:Wp_Baskaran} and then the emergence of conserved flux operators becomes transparent. 

The model can be written as a sum of three terms
of $H_{\text{Kit}}$ as follows
\begin{equation}
H_{\text{Kit}}=H_{\text{Kit}}^{\sigma}+H_{\text{Kit}}^{\sigma T}+H_{\text{Kit}}^{\sigma,\sigma T},\label{eq:H_Kit_separation}
\end{equation}
each of which still preserves a $Z_{2}$ flux structure 
\begin{subequations}\label{eq:HKitAll}
\begin{align}
H_{\text{Kit}}^{\sigma} & =\frac{1}{4}\sum_{\langle ij\rangle_{\gamma}}J_{\gamma}\sigma_{i}^{\gamma}\sigma_{j}^{\gamma},\label{eq:HKitSigma}\\
H_{\text{Kit}}^{\sigma T} & =\sum_{\langle ij\rangle_{\gamma}}J_{\gamma}\sigma_{i}^{\gamma}\sigma_{j}^{\gamma}T_{i}^{\alpha\beta}T_{j}^{\alpha\beta},\label{eq:HKitSigT}\\
H_{\text{Kit}}^{\sigma,\sigma T} & =\frac{1}{2}\sum_{\langle ij\rangle_{\gamma}}J_{\gamma}\sigma_{i}^{\gamma}\sigma_{j}^{\gamma}\left(T_{i}^{\alpha\beta}+T_{j}^{\alpha\beta}\right).\label{eq:HKitSTS}
\end{align}
\end{subequations} 

\subsection{New (soluble) KK models}
Before embarking on a study of the full $S=3/2$ KHM we discuss the individual Hamiltonians of Eq.~\eqref{eq:HKitAll}, two of which turn out to be individually exactly soluble. 

\begin{figure}
\begin{centering}
\includegraphics[width=0.5\columnwidth]{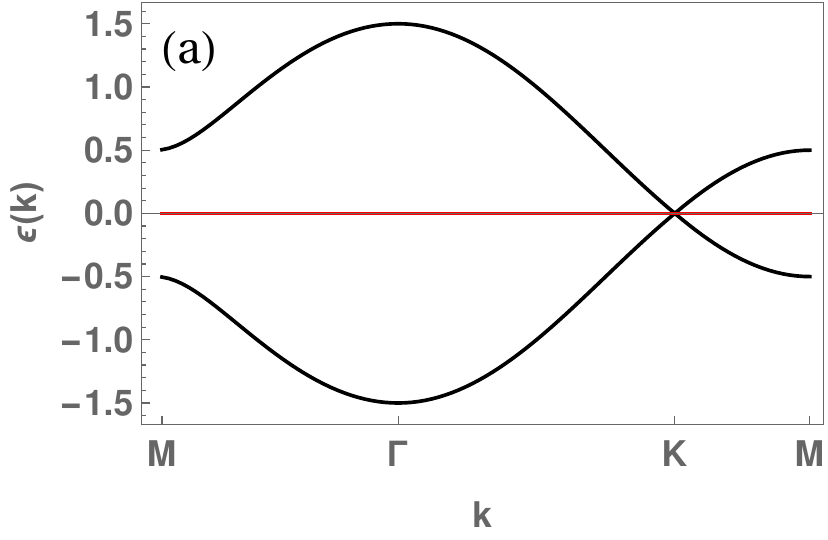}\includegraphics[width=0.5\columnwidth]{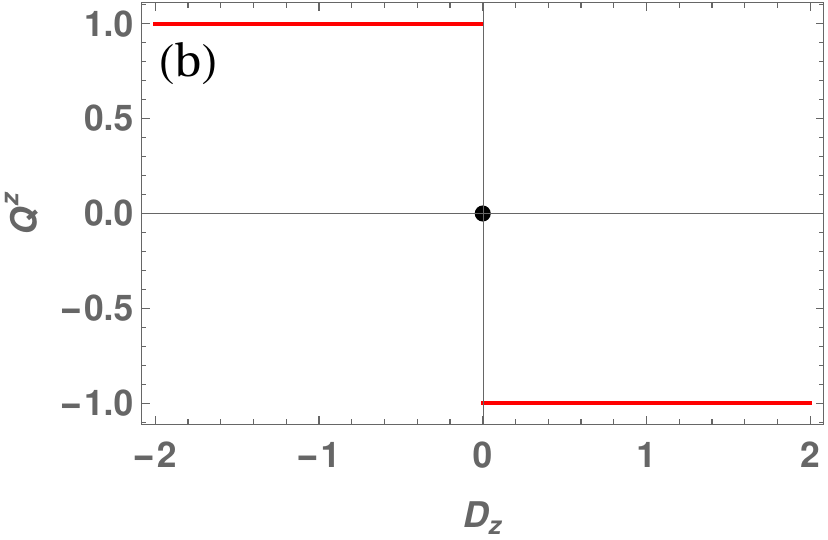}
\par\end{centering}
\caption{\label{fig:Fig2}(a) Dispersion of $H_{\text{Kit}}^{\sigma}$ with flat bands, (b) Strong first-order quantum phase transition induced by the onset of [001] single-ion anisotropy quantified by $D_{z}$.}
\end{figure}

First, we focus on $H_{\text{Kit}}^{\sigma}$ which can be integrated using
the JWT expressed in Eq.~\eqref{eq:XiChiFermion} and corresponds
to the $S=3/2$ exactly solvable model discussed by Bhaskaran \emph{et
al.}~\citep{Baskaran2008}. Its eigenstates $\left|\psi\right\rangle$
can be written as $\left|\psi\right\rangle =\left|\Psi_{\sigma}\right\rangle \otimes\left|\psi_{\mathbf{T}}\right\rangle $,
where $\left|\Psi_{\sigma}\right\rangle $ is an eigenstate of the $S=1/2$ KHM in terms of $\boldsymbol{\sigma}$ operators and $\left|\psi_{\mathbf{T}}\right\rangle$ is an \emph{arbitrary} pseudo-orbital state. Hence, all eigenstates of $H_{\text{Kit}}^{\sigma}$ are $2^{2N}$-fold degenerate, in which $N$ is the number of unit cells. The excitations related to $\left|\Psi_{\sigma}\right\rangle $ at a fixed flux sector correspond to Majorana fermions $\xi$ in Eq.~\eqref{eq:XiChiFermion} with the dispersion of Eq.~\eqref{eq:mu_k}. The arbitrariness of orbital states leads to extra zero-energy flat bands for any choice of exchange couplings and fluxes; Fig.~\ref{fig:Fig2}(a) exemplifies this for the isotropic KHM in the zero-flux sector.

Flat bands are sensitive to small perturbations, and this
can be readily identified in $H_{\text{Kit}}^{\sigma}$. The simplest of this perturbations is the {[}001{]} SIA given by 
\begin{equation}
H_{\text{SIA}}^{z}=D_{z}\sum_{i}\left(S_{i}^{z}\right)^{2}=D_{z}\sum_{i}T_{i}^{z}+\text{const.}\label{eq:z_SIA}
\end{equation}
that commutes with both $W_{p}^{\sigma}$ and $H_{\text{Kit}}^{\sigma}$.
The onset of $D_{z}$ lifts the $S=3/2$ degeneracy by separating
the $S^{z}=\pm3/2$ from the $S^{z}=\pm1/2$,
turning the $H_{\text{Kit}}^{\sigma}$ into a direct sum of two $S=1/2$ KHMs separated by a total energy $2N\left|D_{z}\right|$, each of them characterized by a fixed value of $T^{z}$. The $H_{\text{Kit}}^{\sigma}$
ground state then develops an expectation value $\left\langle T^{z}\right\rangle =+1$
$\left(\left\langle T^{z}\right\rangle =-1\right)$ for infinitesimal
values of $D_{z}<0$ $\left(D_{z}>0\right)$ as indicated in Fig.~\ref{fig:Fig2}(b). More generally, the $\left|D_{z}\right|\rightarrow\infty$ limit of the spin operators reads
\begin{align}
\lim_{\left|D_{z}\right|\rightarrow\infty}\mathbf{S} & _{i}\rightarrow\begin{cases}
\left(-\sigma_{i}^{x},-\sigma_{i}^{y},\frac{\sigma_{i}^{z}}{2}\right), & D_{z}>0,\\
\left(0,0,-\frac{3\sigma_{i}^{z}}{2}\right), & D_{z}<0.
\end{cases}\label{eq:S12_projection}
\end{align}
Thus, large positive values of $D_{z}$ map the $S=3/2$ KHM into its $S=1/2$ version with renormalized coupling constant $J_{z}\rightarrow J_{z}/4$, while large negative $D_{z}$ rapidly maps it into the $S=1/2$ gapped KHM. In other words, the {[}001{]} SIA provides a natural mapping between the $S=3/2$ and $S=1/2$ KHMs while also elucidating the relevance of the $\left\langle T^{z}\right\rangle$ quadrupolar field.

Second, we study the KK model $H_{\text{Kit}}^{\sigma T}$  which turns out to be an exactly solvable model within the class given by Eq.~\eqref{eq:general_flavor_honeycomb}.
This becomes transparent when using the following equivalence between $\Gamma$ matrices and the spin-orbital operators $\left(\boldsymbol{\sigma},\mathbf{T}\right)$
\begin{align}
\Gamma^{1}= & \frac{\sqrt{3}}{3}\left\{ S^{y},S^{y}\right\} =-\sigma^{x}T^{y},\nonumber \\
\Gamma^{2}= & \frac{\sqrt{3}}{3}\left\{ S^{z},S^{x}\right\} =-\sigma^{y}T^{y},\nonumber \\
\Gamma^{3}= & \frac{\sqrt{3}}{3}\left\{ S^{x},S^{y}\right\} =-\sigma^{z}T^{y},\nonumber \\
\Gamma^{4}= & T^{x},\nonumber \\
\Gamma^{5}= & T^{z},\label{eq:sigma_Ty}
\end{align}
by which one can re-expresses Eq.~\eqref{eq:general_flavor_honeycomb}
as 
\begin{equation}
H=\sum_{\langle ij\rangle_{\gamma}}\sum_{a,b=x,y,z}J_{\gamma}^{ab}\sigma_{i}^{\gamma}\sigma_{j}^{\gamma}T_{i}^{a}T_{j}^{b}.
\end{equation}
We note that a related but different soluble KK model has been introduced and studied in Ref.~\citep{Chulliparambil2021}. We will discuss the properties of the exact solution of this new model in the next section in terms of an SO(6) Majorana parton representation of the $S=3/2$ operators~\citep{jin2022unveiling}.

Third, the last model $H_{\text{Kit}}^{\sigma,\sigma T}$ shares the gauge structure of the other two models but within a given flux sector the remaining Majorana problem is still quartic, and thus, not exactly soluble. 

\subsection{Relation between spin-orbital operators and SO(6) Majorana partons}

In Ref.~\citep{jin2022unveiling}, we used an SO(6) Majorana parton representation of the $S=3/2$ operators which allowed us to uncover the static Z$_2$ gauge field description of the flux operators. Here, we will clarify the connection with the pseudo-dipole and pseudo-orbital operators, which  can be written in terms of SO(6) Majorana partons as follows~\citep{Wang2009,Yao2009,Nussinov2009,Yao2011,Chua2011,Farias2020,Chulliparambil2020}
\begin{align}
\boldsymbol{\sigma}_{i}=-\frac{i}{2}\boldsymbol{\eta}_{i}\times\boldsymbol{\eta}_{i}, & \,\mathbf{T}_{i} =-\frac{i}{2}\boldsymbol{\theta}_{i}\times\boldsymbol{\theta}_{i},\nonumber\\ \sigma_{i}^{\alpha}T_{i}^{\beta} &=-i\eta_{i}^{\alpha}\theta_{i}^{\beta},\label{eq:Majorana_parton}
\end{align}
in which $\alpha,\beta=x,y,z$ and $\boldsymbol{\eta}_{i},\boldsymbol{\theta}_{i}$
satisfy 
\begin{align}
\left\{ \eta_{i}^{\alpha},\eta_{j}^{\beta}\right\}  & =\left\{ \theta_{i}^{\alpha},\theta_{j}^{\beta}\right\} =2\delta_{ij}\delta^{\alpha\beta},\nonumber \\
\left\{ \eta_{i}^{\alpha},\theta_{j}^{\beta}\right\}  & =0.
\end{align}

The constraint to the physical Hilbert space is identified by noticing that Eq.~\eqref{eq:sigma_Ty} requires that $\Gamma_{i}^{1}\Gamma_{i}^{2}\Gamma_{i}^{3}\Gamma_{i}^{4}\Gamma_{i}^{5}=-\mathbb{I}$ at all sites. In terms of Eq.~\eqref{eq:Majorana_parton}, the left-hand
side of the equation defines the operator $D_{i}$ given by
\begin{equation}
D_{i}=i\eta_{i}^{\alpha}\eta_{i}^{\beta}\eta_{i}^{\gamma}\theta_{i}^{\alpha}\theta_{i}^{\beta}\theta_{i}^{\gamma}.\label{eq:projector}
\end{equation}
We then demand that a physical state satisfies $D_{i}=1,\forall i$. Equivalently, we can formally write a projector operator $P$~\citep{YaoPRL2007}
\begin{equation}
P=\prod_{i}\frac{1+D_{i}}{2}\equiv P^{\prime}\left(\frac{1+D}{2}\right),\label{eq:projector2}
\end{equation}
in which $D=\prod_{i=1}^{2N}D_{i}$ and $P^{\prime}$ is the sum over all inequivalent gauge transformations. A physical state $\left|\psi\right\rangle$
is considered physical if, and only if, $\left|\psi\right\rangle =P\left|\psi\right\rangle $.
An explicit formula for $D$ can be derived following Refs.~\cite{Pedrocchi2011,Zschocke2015} and is given in Appendix~\ref{sec:Projection} for SO(6) Majorana fermions.

\begin{figure*}
\begin{centering}
\includegraphics[width=0.5\columnwidth]{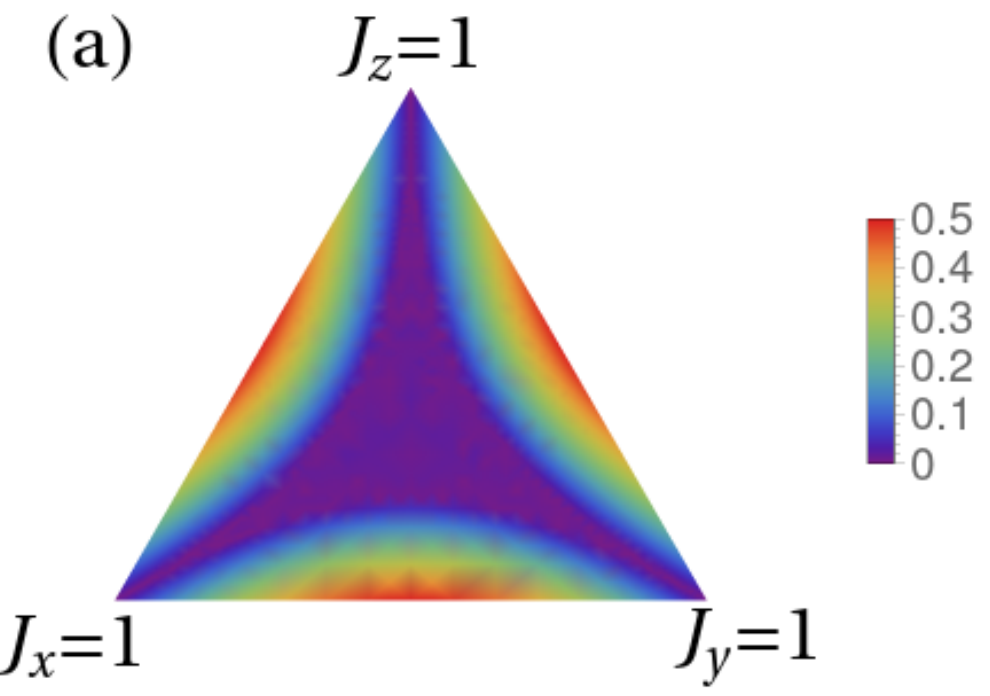}\includegraphics[width=0.5\columnwidth]{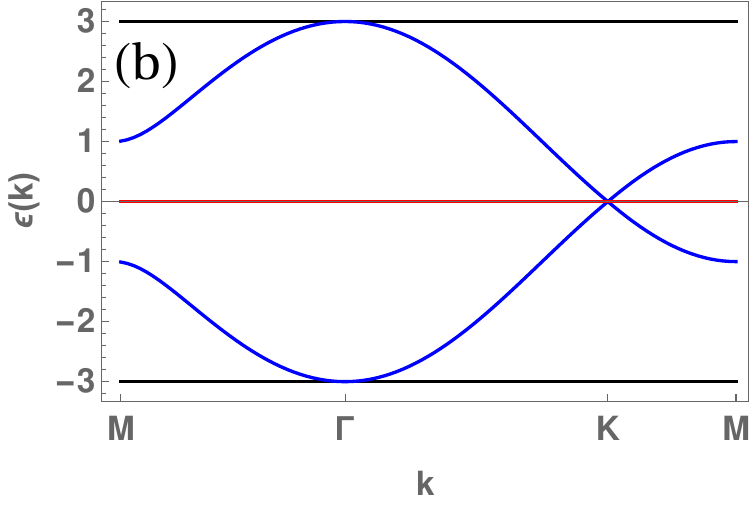}\includegraphics[width=0.5\columnwidth]{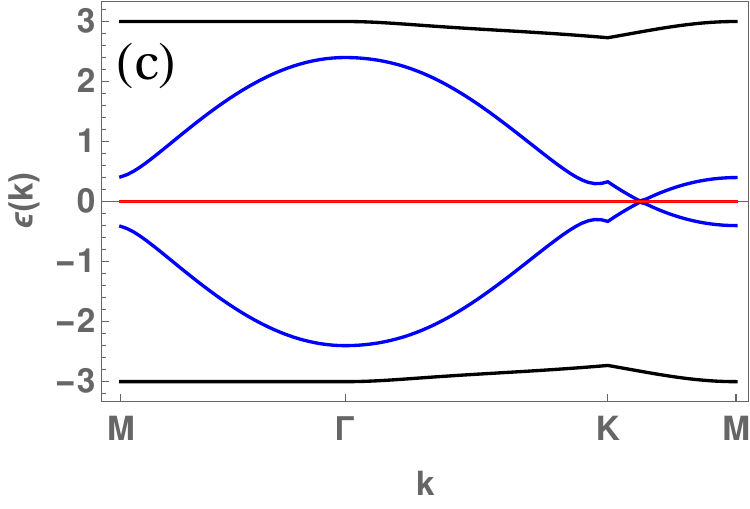}\includegraphics[width=0.5\columnwidth]{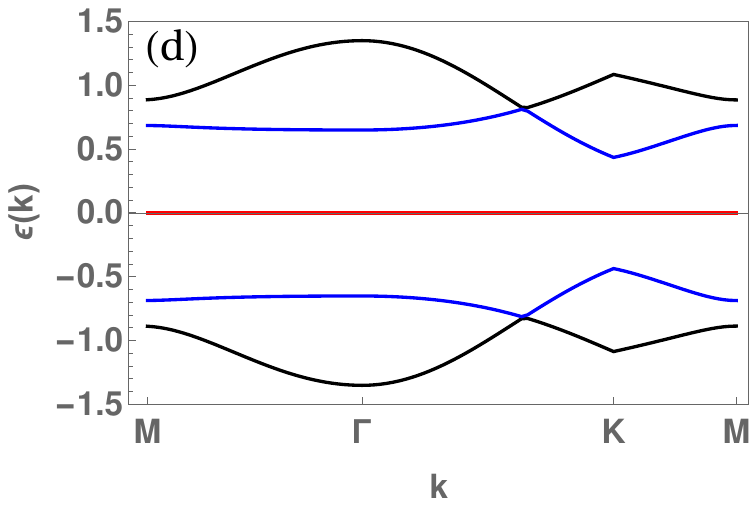}
\par\end{centering}
\caption{\label{fig:Phase-diagram-St} (a) Phase diagram of the spin-orbital model $H_{\text{Kit}}^{\sigma T}$ on the plane $J_{x}+J_{y}+J_{z}=1$ with positive coupling constants,
in which the dark blue area corresponds to gapless phases. The graphics (b-d) correspond to the spectrum of excitations along high-symmetry lines with the following coupling constants (b) $J_{x}=J_{y}=J_{z}=1$,
(c) $J_{x}=J_{y}=1.0$, $J_{z}=0.7$, and (d) $J_{y}=0.1$, $J_{x}=J_{z}=0.45$.}
\end{figure*}

We can now use the partons for an exact solution of the model in Eq. (\ref{eq:HKitSigT}) as it reads in the new form
\begin{equation}
H_{\text{Kit}}^{\sigma T}=\sum_{\left\langle ij\right\rangle _{\gamma}}J_{\gamma}\hat{u}_{\left\langle ij\right\rangle _{\gamma}}i\theta_{i}^{\alpha\beta}\theta_{j}^{\alpha\beta},\label{eq:HsigTparton}
\end{equation}
where $\hat{u}_{\langle ij\rangle_{\gamma}}$ is the same $Z_{2}$
bond operator defined for the $S=1/2$ KHM and $\theta^{xy}=\theta^{z}$,
$\theta^{yz(zx)}=\left(-\theta^{z}\pm\sqrt{3}\theta^{x}\right)/2$.
Notice that $\theta^{y}$ fermions are absent and lead to zero-energy flat bands at any flux sector or choice of exchange couplings, in analogy to $H_{\text{Kit}}^{\sigma}$.
The ground state is again in the zero-flux sector \citep{Lieb1994}, for which the dispersive bands can be gapped or gapless according to the values of $J_{\gamma}$ (see Fig.~\ref{fig:Phase-diagram-St}).
The isotropic case in Fig.~\ref{fig:Phase-diagram-St}(b) displays a band whose dispersion is exactly $\epsilon(\mathbf{k})$ in
Eq.~\eqref{eq:mu_k}, i.e., it is formally the same as the original Kitaev model. This band is sandwiched between two flat bands with energy given exactly by $E=0$ and $E=3J$. Away from the isotropic limit, the high-energy flat band acquires a dispersion and the intermediate
bands deviate from $\epsilon (\mathbf{k})$, see Figs. \ref{fig:Phase-diagram-St}(c)
and (d). 

Finally, we would like to point out a few interesting aspects in relation to the exact solution of the first Hamiltonian $H_{\text{Kit}}^{\sigma}$ in terms of SO(6) Majorana fermions. Although a solution for this model can be obtained directly through the SO(3) representation of $\boldsymbol{\sigma}$ given by Eq.~\eqref{eq:Majorana_parton}~\citep{Fu2018},
it is instructive to obtain an alternative representation of the pseudo-dipoles through the $D_{i}$ operators~\citep{Schaden2022arXiv}. By evaluating $D_{i}\boldsymbol{\sigma}_{i}$
and then setting $D_{i}=1$, we obtain $\sigma_{i}^{\gamma}=-i\eta_{i}^{\gamma}\theta_{i}^{0}$,
in which 
\begin{subequations} 
\begin{equation}
\theta_{i}^{0}\equiv-i\theta_{i}^{x}\theta_{i}^{y}\theta_{i}^{z}=-i\theta_{i}^{\alpha}\theta_{i}^{\beta}\theta_{i}^{\gamma}.\label{eq:theta_0}
\end{equation}
\end{subequations} 
The expression for the pseudo-dipoles is the same as the one expressed for the $S=1/2$ KHM in Eq.~\eqref{eq:theta_0_parton}, but the fact that $\theta_{i}^{0}$ is now a product of three Majorana flavors demands more careful analysis. $\theta_{i}^{0}$ satisfies
the Majorana fermion algebra $\left\{ \theta_{i}^{0},\theta_{j}^{0}\right\} =2\delta_{ij}$
and $\left\{ \theta_{i}^{0},\eta_{j}^{\gamma}\right\} =0$, so that the spectrum of matter excitations of $H_{\text{Kit}}^{\sigma}$ can still be exactly known by mapping the Hamiltonian to a free fermion-like problem. However, the dimension of the $\theta_{i}^{0}$ Hilbert space is twice that of a conventional Majorana fermion, which reflects the independence of $H_{\text{Kit}}^{\sigma}$ in relation to orbital states. $\theta_{i}^{0}$
is also very sensitive to local orbital operators such as the SIA in Eq.~\eqref{eq:z_SIA}, which is represented like
\begin{equation}
H_{\text{SIA}}^{z}=-\sum_{j}D_{z}i\theta_{j}^{x}\theta_{j}^{y}.\label{eq:z-SIA-parton}
\end{equation}
Combining Eq.~\eqref{eq:theta_0} and Eq.~\eqref{eq:z-SIA-parton}, we observe that the SIA along the $z$-direction ``freezes'' the Majorana flavors $\theta^{x}$ and $\theta^{y}$ and allows the replacement
$\theta_{i}^{0}\rightarrow-\text{sign}\left(D_{z}\right)\theta_{i}^{z}$ in accordance with our previous discussion. 

Remarkably, although $H_{\text{Kit}}^{\sigma}$ and $H_{\text{Kit}}^{\sigma T}$ are individually exactly soluble, their sum is not due to the same site commutation
\begin{equation}
\left[\theta_{i}^{0},\theta_{i}^{\gamma}\right]=0.\label{eq:theta_0_comm}
\end{equation}
Thus, the set of four operators $\theta_{i}^{0},\theta_{i}^{x,y,z}$
do not behave as mutual Majorana fermions when all present, but instead are operators akin to what is known in the literature as \emph{Greenberg parafermions}~\citep{Green1953,GreenbergMessiah1965,Macfarlane1994}. Returning to the JWT expressed in Eq.~\eqref{eq:XiChiFermion}, it is possible to demonstrate an equivalence between $\theta_{i}^{0}$ and $\xi_{i}$, as well as between $\theta_{i}^{\gamma}$ and $\xi_{i}T_{i}^{\gamma}$
(the interested reader can follow Appendix~\ref{sec:JWT_S32}). Besides giving an interpretation to the SO(6) Majorana partons in terms of strings of operators this observation could possibly be useful for more general classes of parafermions in $S=3/2$
models~\citep{Vaezi2014,Barkeshli2015,Alicea2016,Fendley2014}.

\section{Parton Mean-Field Theory of the $S=3/2$ KHM \label{sec:Parton-MFT}}

After having discussed the different representations of $S=3/2$ operators in terms of spin-orbital operators and SO(6) partons, we would like to study the full $S=3/2$ KHM that is explicitly given by
\begin{align}
H_{\text{Kit}} & =\sum_{\left\langle ij\right\rangle _{\gamma}}J_{\gamma}\hat{u}_{\left\langle ij\right\rangle _{\gamma}}i\theta_{i}^{\alpha\beta}\theta_{j}^{\alpha\beta}\nonumber \\
 & +\sum_{\left\langle ij\right\rangle _{\gamma}}\frac{J_{\gamma}}{4}\hat{u}_{\left\langle ij\right\rangle _{\gamma}}i\theta_{i}^{0}\theta_{j}^{0}\nonumber \\
 & +\sum_{\left\langle ij\right\rangle _{\gamma}}\frac{J_{\gamma}}{2}\hat{u}_{\left\langle ij\right\rangle _{\gamma}}\left(i\theta_{i}^{0}\theta_{j}^{\alpha\beta}+i\theta_{i}^{\alpha\beta}\theta_{j}^{0}\right).\label{eq:HKit_full_parton}
\end{align}

We emphasize that the first line of Eq.~ \ref{eq:HKit_full_parton} is quadratic in terms of SO(6) Majorana fermions, whereas the second line is sextic, and the third is quartic. In order to proceed with analytical calculations, we need to perform a mean-field decoupling in terms of the following parameters~\citep{jin2022unveiling}
\begin{align}
Q_{i}^{\gamma} & =\left\langle T_{i}^{\gamma}\right\rangle =-\left\langle i\theta_{i}^{\alpha}\theta_{i}^{\beta}\right\rangle ,\nonumber \\
\Delta_{ij}^{\lambda\mu} & =-\left\langle i\theta_{i}^{\lambda}\theta_{j}^{\mu}\right\rangle ,\label{eq:order_parameters}
\end{align}
in which $i$ and $j$ are nearest-neighbor sites and the averages are obtained self-consistently. More explicitly, we write
\begin{align}
i\theta_{i}^{0}\theta_{j}^{\alpha\beta}+i\theta_{i}^{\alpha\beta}\theta_{j}^{0} & \approx\sum_{p=x,y,z}\left(Q_{i}^{p}i\theta_{i}^{p}\theta_{j}^{\alpha\beta}+\Delta_{\left\langle ij\right\rangle _{\gamma}}^{p,\alpha\beta}i\theta_{i}^{q}\theta_{i}^{r}\right)\nonumber \\
 & +\sum_{p=x,y,z}\left(Q_{j}^{p}i\theta_{i}^{\alpha\beta}\theta_{j}^{p}+\Delta_{\left\langle ij\right\rangle _{\gamma}}^{\alpha\beta,p}i\theta_{i}^{q}\theta_{i}^{r}\right),
\end{align}
\begin{align}
i\theta_{i}^{0}\theta_{j}^{0}\approx & -\sum_{a=x,y,z}\left\langle \theta_{i}^{c}\theta_{j}^{x}\theta_{j}^{y}\theta_{j}^{z}\right\rangle i\theta_{i}^{a}\theta_{i}^{b}\nonumber \\
 & -\sum_{a=x,y,z}\left\langle \theta_{i}^{x}\theta_{i}^{y}\theta_{i}^{z}\theta_{j}^{c}\right\rangle i\theta_{j}^{a}\theta_{j}^{b}\nonumber \\
 & -\sum_{a,a^{\prime}=x,y,z}\left\langle \theta_{i}^{b}\theta_{i}^{c}\theta_{j}^{b^{\prime}}\theta_{j}^{c^{\prime}}\right\rangle i\theta_{i}^{a}\theta_{j}^{a^{\prime}}. \label{eq:MFT_sextic}
\end{align}
The quartic averages in Eq.~\eqref{eq:MFT_sextic} are written in terms of Eq.\eqref{eq:order_parameters} using Wick's theorem, which states 
\begin{align}
-\left\langle \theta_{i}^{c}\theta_{j}^{x}\theta_{j}^{y}\theta_{j}^{z}\right\rangle  & =\Delta_{\left\langle ij\right\rangle }^{cx}Q_{j}^{x}+\Delta_{\left\langle ij\right\rangle }^{cy}Q_{j}^{y}+\Delta_{\left\langle ij\right\rangle }^{cz}Q_{j}^{z},\nonumber\\
-\left\langle \theta_{i}^{x}\theta_{i}^{y}\theta_{i}^{z}\theta_{j}^{c}\right\rangle  & =Q_{i}^{x}\Delta_{\left\langle ij\right\rangle }^{xc}+Q_{i}^{y}\Delta_{\left\langle ij\right\rangle }^{yc}+Q_{i}^{z}\Delta_{\left\langle ij\right\rangle }^{zc},\\
-\left\langle \theta_{i}^{b}\theta_{i}^{c}\theta_{j}^{b^{\prime}}\theta_{j}^{c^{\prime}}\right\rangle  & =Q_{i}^{a}Q_{j}^{a^{\prime}}-\Delta_{\left\langle ij\right\rangle }^{bb^{\prime}}\Delta_{\left\langle ij\right\rangle }^{cc^{\prime}}+\Delta_{\left\langle ij\right\rangle }^{bc^{\prime}}\Delta_{\left\langle ij\right\rangle }^{cb^{\prime}}.\nonumber\label{eq:Th0_decoupling}
\end{align}

Although a large number of mean-field parameters are introduced in Eq.~\eqref{eq:order_parameters}, a closer analysis of $Q_{i}^{\gamma}$ and $\Delta_{ij}^{\lambda\mu}$ shows that many of them vanish or are related by at most a negative sign factor, evincing symmetry constraints. In the following, we revisit the $S=3/2$ KSL under the assumption that it preserves TRS and spatial symmetries, which greatly reduces the number of independent PMFT parameters. We keep our analysis for exchange parameters along the line $J_{x}=J_{y}=1$ and $D_{z}=0$, for which
it displays one mirror symmetry $M^{b}$, whose mirror operator lies along the $a$ axis, and a $\pi$ rotation around the $b$ axis (see Fig. \ref{fig:mirrors}). Whenever $Z_{2}$ gauge operators were fixed, we assume that the operation was performed in the zero-flux sector. We also analyze the $C_{3}$ rotation symmetry around the $c$ axis
on the isotropic model, which is crucial to understand the strong first-order quantum phase transition that separates the distinct KSL phases.

\subsection{Symmetries of the $S=3/2$ KHM }

\subsubsection{Time-reversal symmetry}

Due to the oddness of spin under time-reversal $\mathcal{T}$, Eq. (\ref{eq:orbitals_as_spins}) implies that the pseudo-dipoles and pseudo-orbitals transform like
\begin{align}
\mathcal{T}\boldsymbol{\sigma}_{i}\mathcal{T}^{-1} & =-\boldsymbol{\sigma}_{i},\nonumber \\
\mathcal{T}\mathbf{T}_{i}\mathcal{T}^{-1} & =\left(T_{i}^{x},-T_{i}^{y},T_{i}^{z}\right).
\end{align}
By including the effect of complex conjugation $\mathcal{T}i\mathcal{T}^{-1}=-i$, the corresponding action of $\mathcal{T}$ on the SO(6) Majorana partons is~\citep{Natori2016}
\begin{align}
\mathcal{T}\boldsymbol{\eta}_{i}\mathcal{T}^{-1} & =\left(\eta_{i}^{x},\eta_{i}^{y},\eta_{i}^{z}\right),\nonumber \\
\mathcal{T}\boldsymbol{\theta}_{i}\mathcal{T}^{-1} & =\left(\theta_{i}^{x},-\theta_{i}^{y},\theta_{i}^{z}\right),\label{eq:TR_fermions}
\end{align}
upon which we define the indices $\mathfrak{t}_{x}=\mathfrak{t}_{z}=1$,
$\mathfrak{t}_{y}=-1$ for the matter fermions. The transformation of products of order parameters and $Z_{2}$ gauge variables is then given by
\begin{equation}
\mathcal{T}\left(i\hat{u}_{ij}^{\gamma}\theta_{i}^{\lambda}\theta_{j}^{\mu}\right)\mathcal{T}^{-1}=\mathfrak{t}_{\lambda}\mathfrak{t}_{\mu}i\hat{u}_{ij}^{\gamma}\theta_{i}^{\lambda}\theta_{j}^{\mu},\label{eq:TRS_OP}
\end{equation}
where we used $\mathcal{T}\hat{u}_{ij}^{\gamma}\mathcal{T}^{-1}=-\hat{u}_{ij}^{\gamma}$.
Let us then fix the gauge operators. If the ground state  $\left|\psi_{0}\right\rangle $
does not break a symmetry $\mathcal{S}$, then $\left\langle \psi_{0}\left|\mathcal{O}\right|\psi_{0}\right\rangle =\left\langle \psi_{0}\left|\mathcal{S}\mathcal{O}\mathcal{S}^{-1}\right|\psi_{0}\right\rangle $.
Eq.~\eqref{eq:TRS_OP} implies that
when $\mathcal{O}$ is a product of two matter fermions, the parameters must fulfill
\begin{equation}
\Delta_{\left\langle ij\right\rangle _{\gamma}}^{\lambda y}=\Delta_{\left\langle ij\right\rangle _{\gamma}}^{y\lambda}=0\text{ if }\lambda\neq y.\label{eq:TRS_constraint}
\end{equation}
An important consequence of this relation is that, in a time-reversal
symmetric QSL, $\theta^{y}$ hybridizes with other Majorana flavors
only through the onsite order parameters $Q_{i}^{z}=-\left\langle i\theta_{i}^{x}\theta_{i}^{y}\right\rangle $
or $Q_{i}^{x}=-\left\langle i\theta_{i}^{y}\theta_{i}^{z}\right\rangle $. 

\subsubsection{Mirror and $C_{2}$ rotation}

The effect of spatial symmetries on the Kitaev model is more readily
understood in terms of $\left(S^{a},S^{b},S^{c}\right)$ spins in
the crystallographic frame, whose relation to the spins on the cubic
axes is~\citep{Janssen_2019,Consoli2020,Maksimov2020} 
\begin{align}
S^{x} & =\frac{S^{a}}{\sqrt{6}}-\frac{S^{b}}{\sqrt{2}}+\frac{S^{c}}{\sqrt{3}},\nonumber \\
S^{y} & =\frac{S^{a}}{\sqrt{6}}+\frac{S^{b}}{\sqrt{2}}+\frac{S^{c}}{\sqrt{3}},\nonumber \\
S^{z} & =-\sqrt{\frac{2}{3}}S^{a}+\frac{S^{c}}{\sqrt{3}}.
\end{align}
The action of $R=M^{b},C_{2}$ on an isolated spin is $R\left(S^{a},S^{b},S^{c}\right)R^{-1}\equiv R\left(S^{a},S^{b},S^{c}\right)=\left(-S^{a},S^{b},-S^{c}\right)$,
and leads to
\begin{equation}
R\mathbf{S}_{i}=\left(-S_{R(i)}^{y},-S_{R(i)}^{x},-S_{R(i)}^{z}\right).\label{eq:mirror_spins}
\end{equation}
The most relevant difference between $M^{b}$ and $C_{2}$ is that
$i$ and $M^{b}(i)$ are on opposite sublattices, whereas $i$ and
$C_{2}(i)$ are on the same. The application of Eq.~\eqref{eq:mirror_spins}
in Eq.~\eqref{eq:orbitals_as_spins} implies that
\begin{align}
RT_{i}^{z} & =T_{R(i)}^{z},\nonumber \\
RT_{i}^{x} & =-T_{R(i)}^{x},\nonumber \\
RT_{i}^{y} & =-T_{R(i)}^{y}.\label{eq:mirror_orbitals-My}
\end{align}
Therefore, if $C_{2}$ and translation symmetries are preserved,
\begin{equation}
Q_{X}^{x/y}=-Q_{X}^{x/y}\implies Q_{X}^{x/y}=0.
\end{equation}
Hence, the only onsite order parameter allowed by spatial symmetries
is $Q^{z}$. 

\begin{figure}
\begin{centering}
\includegraphics[width=0.6\columnwidth]{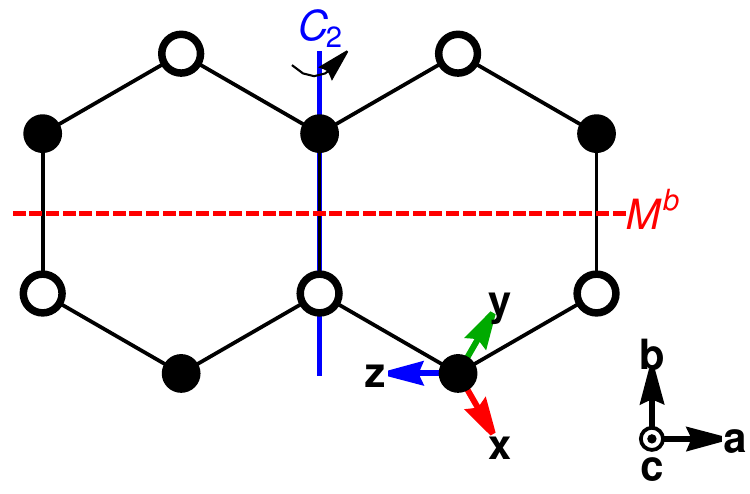}
\par\end{centering}
\caption{\label{fig:mirrors} Two plaquettes of the honeycomb lattice. The
figure displays (i) the crystallographic axes $(a,b,c)$, (ii) the
projection of the $(x,y,z)$ axes onto the $ab$ plane, (iii) the
mirror elements $M^{b}$ and the $C_{2}$ rotation axis, and (iv)
the distinction between even and odd sublattices.}
\end{figure}

To evaluate the effect of symmetry operators over $\Delta_{ij}^{\lambda\mu}$,
we first observe that Eq.~\eqref{eq:mirror_spins} implies that 
\begin{align}
R\boldsymbol{\eta}_{i} & =\left(-\eta_{R(i)}^{y},-\eta_{R(i)}^{x},-\eta_{R(i)}^{z}\right),\nonumber \\
R\boldsymbol{\theta}_{i} & =\left(-\theta_{R(i)}^{x},-\theta_{R(i)}^{y},\theta_{R(i)}^{z}\right).\label{eq:mirror_fermions}
\end{align}
After defining
\begin{equation}
s_{\lambda}=\begin{cases}
-1, & \text{if }\lambda=x,y\\
1 & \text{if }\lambda=z,
\end{cases}\label{eq:s_index}
\end{equation}
Eq.~\eqref{eq:mirror_fermions} yields 
\begin{align}
M^{b}\left(i\hat{u}_{ij}^{\gamma}\theta_{i}^{\lambda}\theta_{j}^{\mu}\right) & =s_{\lambda}s_{\mu}i\hat{u}_{M_{j}^{b}M_{i}^{b}}^{m(\gamma)}\theta_{M_{j}^{b}}^{\mu}\theta_{M_{i}^{b}}^{\lambda},\nonumber \\
C_{2}\left(i\hat{u}_{ij}^{\gamma}\theta_{i}^{\lambda}\theta_{j}^{\mu}\right) & =s_{\lambda}s_{\mu}i\hat{u}_{C_{2}\left(i\right)C_{2}\left(j\right)}^{m(\gamma)}\theta_{C_{2}\left(i\right)}^{\lambda}\theta_{C_{2}\left(j\right)}^{\mu},
\end{align}
in which $m(x)=y$, $m(y)=x$, and $m(z)=z$ are indices related to
the bond transformation under $R$. The $C_{2}$ symmetry of the Hamiltonian
then implies $\Delta_{\left\langle ij\right\rangle _{\gamma}}^{\lambda\mu}=s_{\lambda}s_{\mu}\Delta_{\left\langle ij\right\rangle _{m(\gamma)}}^{\lambda\mu},$
which leads to
\begin{align}
\Delta_{\left\langle ij\right\rangle _{z}}^{zx} & =\Delta_{\left\langle ij\right\rangle _{z}}^{xz}=\Delta_{\left\langle ij\right\rangle _{z}}^{zy}=\Delta_{\left\langle ij\right\rangle _{z}}^{yz}=0,\nonumber \\
\Delta_{\left\langle ij\right\rangle _{y}}^{\lambda\mu} & =s_{\lambda}s_{\mu}\Delta_{\left\langle ij\right\rangle _{x}}^{\lambda\mu}.\label{eq:C2_constraints}
\end{align}
Applying a similar reasoning to $M^{b}$, we find 
\[
\Delta_{\left\langle ij\right\rangle _{y}}^{\lambda\mu}=s_{\lambda}s_{\mu}\Delta_{\left\langle ij\right\rangle _{x}}^{\mu\lambda},
\]
which in combination with Eq.~\eqref{eq:C2_constraints} give
\begin{align}
\Delta_{\left\langle ij\right\rangle _{x}}^{\lambda\mu} & =\Delta_{\left\langle ij\right\rangle _{x}}^{\mu\lambda},\,\Delta_{\left\langle ij\right\rangle _{y}}^{\lambda\mu}=\Delta_{\left\langle ij\right\rangle _{y}}^{\mu\lambda}.
\end{align}
The results gathered in this section imply that $\Delta_{\left\langle ij\right\rangle _{x}}^{zx}$
is the only non-zero mixed-flavor order parameter $\Delta$, and all others either vanish or are related to it by symmetry. We also confirmed this constraint numerically along the line $J_{x}=J_{y}$. 

\subsubsection{$C_{3}$ symmetry}

The isotropic point is a critical point of strong first-order phase
transitions~\citep{jin2022unveiling} which motivates a closer look. The key symmetry distinction of the KHM in this point to others
discussed above is its invariance under $C_{3}$ rotations, whose
effect on spins is given by
\begin{equation}
C_{3}\left(\begin{array}{c}
S_{\mathbf{r}X}^{x}\\
S_{\mathbf{r}X}^{y}\\
S_{\mathbf{r}X}^{z}
\end{array}\right)=\left(\begin{array}{c}
S_{\left(R_{3}\mathbf{r}\right)X}^{y}\\
S_{\left(R_{3}\mathbf{r}\right)X}^{z}\\
S_{\left(R_{3}\mathbf{r}\right)X}^{x}
\end{array}\right),
\end{equation}
in which we see that the sublattices remain invariant under rotation.
The corresponding parton transformations are 
\begin{align}
C_{3}\left(\begin{array}{c}
\eta_{\mathbf{r}X}^{x}\\
\eta_{\mathbf{r}X}^{y}\\
\eta_{\mathbf{r}X}^{z}
\end{array}\right) & =\left(\begin{array}{c}
\eta_{\left(C_{3}\mathbf{r}\right)X}^{y}\\
\eta_{\left(C_{3}\mathbf{r}\right)X}^{z}\\
\eta_{\left(C_{3}\mathbf{r}\right)X}^{x}
\end{array}\right),\nonumber \\
C_{3}\left(\begin{array}{c}
\theta_{\mathbf{r}X}^{x}\\
\theta_{\mathbf{r}X}^{y}\\
\theta_{\mathbf{r}X}^{z}
\end{array}\right) & =\left(\begin{array}{ccc}
-\frac{1}{2} & 0 & -\frac{\sqrt{3}}{2}\\
0 & 1 & 0\\
\frac{\sqrt{3}}{2} & 0 & -\frac{1}{2}
\end{array}\right)\left(\begin{array}{c}
\theta_{\left(C_{3}\mathbf{r}\right)X}^{x}\\
\theta_{\left(C_{3}\mathbf{r}\right)X}^{y}\\
\theta_{\left(C_{3}\mathbf{r}\right)X}^{z}
\end{array}\right).
\end{align}
These equations are enough to enforce several constraints between
the order parameters that are tabled explicitly in Appendix~\ref{sec:C3_relation_OP}.
In particular, the quadrupolar order parameters satisfy
\begin{align*}
Q_{X}^{z} & =-\frac{1}{2}Q_{X}^{z}+\frac{\sqrt{3}}{2}Q_{X}^{x},\\
Q_{X}^{x} & =-\frac{1}{2}Q_{X}^{x}-\frac{\sqrt{3}}{2}Q_{X}^{z},
\end{align*}
and therefore
\begin{align}
Q_{X}^{z}= & 0.
\end{align}
In other words, if the isotropic KSL does not break symmetries, then
we do not expect any pseudo-orbital order at the isotropic point.
This result is in sharp contrast to the semiclassical QSL proposed
in Ref. \citep{IoannisNatComm2018}, since the kekule pattern of the
dimers impose an order of $Q^{z}$ and $Q^{x}$.

\subsection{Constrained Mean-field Hamiltonian}

The symmetry constrained PMFT parameters for the zero-flux sector can be summarized as follows
\begin{align}
Q^{x}=Q^{y} &=0,\nonumber \\ Q_{A}^{z} &=Q_{B}^{z},\nonumber \\
\Delta_{\left\langle ij\right\rangle _{\gamma}}^{ab} & =\Delta_{\left\langle ij\right\rangle _{\gamma}}^{ba},\nonumber \\
\Delta_{\left\langle ij\right\rangle _{y}}^{ab} & =s_{a}s_{b}\Delta_{\left\langle ij\right\rangle _{x}}^{ab},\nonumber \\
\Delta_{\left\langle ij\right\rangle _{z}}^{xz}=\Delta_{\left\langle ij\right\rangle _{z}}^{zx} & =0,\nonumber \\
\Delta_{\left\langle ij\right\rangle _{\gamma}}^{\lambda y}=\Delta_{\left\langle ij\right\rangle _{\gamma}}^{y\lambda} & =0,\text{ if }\lambda\neq y,
\end{align}
i.e., there are only eight independent, non-vanishing  parameters to be computed self-consistently
\begin{equation}
Q^{z},\Delta_{\left\langle ij\right\rangle _{z}}^{aa},\Delta_{\left\langle ij\right\rangle _{x}}^{aa},\Delta_{\left\langle ij\right\rangle _{x}}^{zx}.\label{eq:non_vanishing_OP}
\end{equation}
For SIA preserving mirror, $C_{2}$, and TRS the results above are valid for $D_{z}\neq0$. At the isotropic point, we find only three non-vanishing and independent parameters given by $\Delta_{\left\langle ij\right\rangle _{z}}^{aa}$. The order parameters obtained through unconstrained PMFT in Ref.~\citep{jin2022unveiling}
are consistent with these results, thus demonstrating that the $S=3/2$ KSLs are the most general $S=3/2$ Majorana QSL preserving all the model's symmetries while minimizing the energy.

\begin{figure*}
\begin{centering}
\includegraphics[width=0.5\columnwidth]{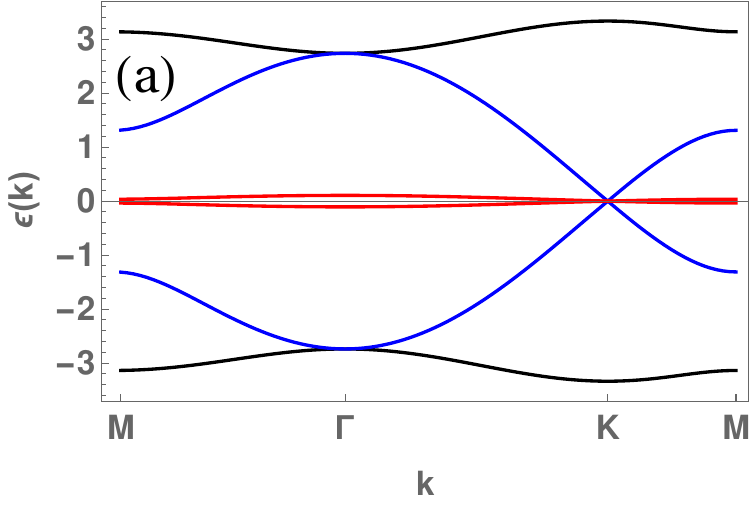}\includegraphics[width=0.5\columnwidth]{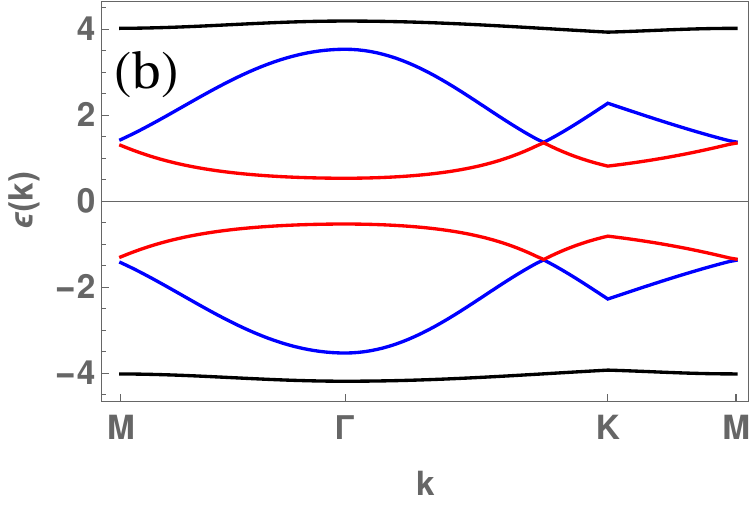}\includegraphics[width=0.5\columnwidth]{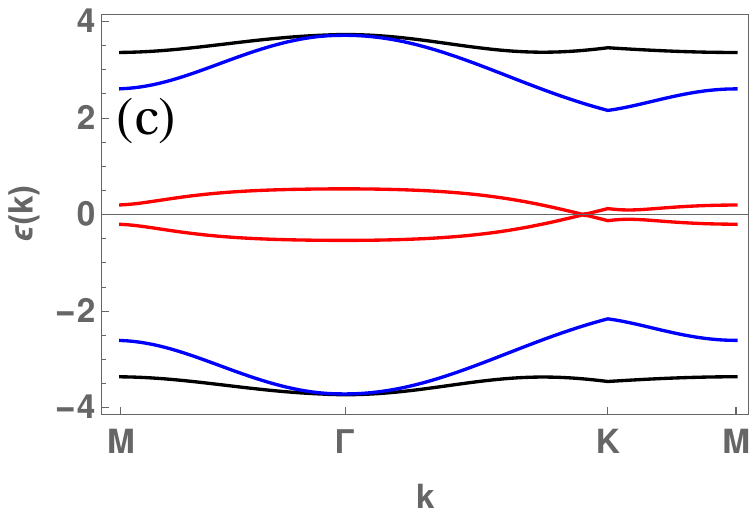}\includegraphics[width=0.5\columnwidth]{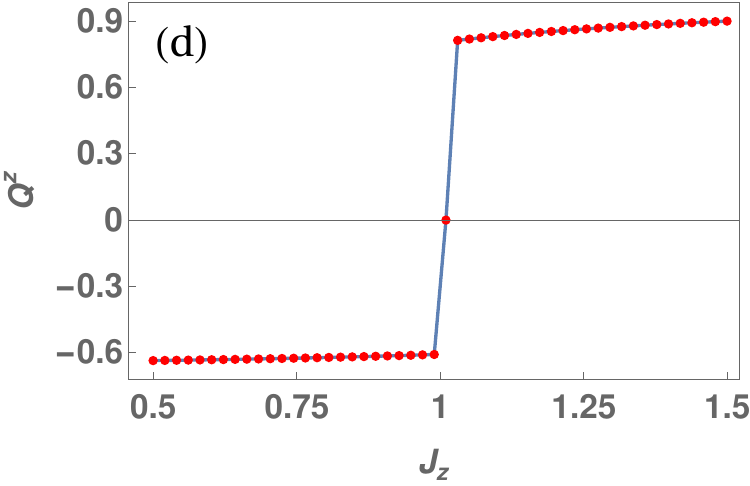}
\par\end{centering}
\caption{\label{fig:Representative_cases_3_2KSL} The graphics (a-c) exemplify the Majorana dispersion of the $S=3/2$ KSL on the (a) isotropic, (b) gapped $J_{z}=1.01$, and (c) gapless $J_{z}=0.99$ cases. Panel (d) shows the evolution of the quadrupolar expectation value $Q^{z}$ as a function of $J_{z}$}
\end{figure*}

We are now ready to give an in-depth description
of the different KSL phases starting with the isotropic case~\citep{jin2022unveiling}, as shown in Fig.~\ref{fig:phasediagramNC}.
$C_{3}$-symmetry constraints enforce that $H_{\text{Kit,MFT}}^{\sigma,\sigma T}=0$,
such that the KHM at this point is described by $H_{\text{Kit}}^{\sigma T}$ 
perturbed by a model whose entries are proportional to $J\left(\Delta_{\left\langle ij\right\rangle _{z}}^{aa}\right)^{2}$ from the six fermion interaction of $H_{\text{Kit}}^{\sigma}$, see Eq.~\eqref{eq:Th0_decoupling}. The qualitative properties of the isotropic KSL can be  understood from the ``parent Hamiltonian'' $H_{\text{Kit}}^{\sigma T}$ but with an interaction induced  small dispersion to the isotropic QSL flat bands and renormalization of the dispersive bands, as can be seen by comparing
Fig.~\ref{fig:Representative_cases_3_2KSL}(a) and Fig.~\ref{fig:Phase-diagram-St}(b).

The symmetry constraint preventing the hybridization of $\theta^{y}$ with mobile $\theta^{x,z}$ fermions only appear at the isotropic point and for $D_{z}=0$. For all other $\left(J_{z},D_{z}\right)$ points, a nonzero $Q^{z}=\left\langle T^{z}\right\rangle$ expectation value appears reducing the energy by strongly affecting the low-energy flat band. 
The presence of the flat band, therefore, explains the strong first-order phase transitions in the neighborhood of the isotropic point. Figs.~\ref{fig:Representative_cases_3_2KSL}(b)
and (c) show that the Majorana fermion dispersion of both the gapped ($J_{z}>1$) and the gapless ($J_{z}<1$) phases are very different from the isotropic one even for small deviations of $J_{z}=1$. Once the transition occurs, Fig.~\ref{fig:Representative_cases_3_2KSL}(d) indicates that $Q^{z}$ varies slowly as a function of $J_{z}$.

Let us now consider the gapped KSL exemplified by those on the line $J_{z}>1,\,D_{z}=0$. A qualitative picture of this KSL is understood by starting from $J_{x}=J_{y}=0$ (or $J_{z}\rightarrow\infty$), which displays a $2^{N}$-fold degenerate ground state composed by all direct products of antiferromagnetic dimers with $S^{z}=\pm3/2$. All states in this manifold are characterized by the same quadrupolar
order $Q^{z}=+1$ at all sites. Introducing small values of $J_{x}$ and $J_{y}$ allows us to derive a toric code model~\citep{Kitaev2006} at the 12th order in perturbation theory for $S=3/2$. The toric-code exchange coupling thus scales as $\left(J_{z}\right)^{-11}$, which implies a rapid decay of the flux gap. This feature is manifest in the DMRG simulations, for which the plaquette operators $W_{p}^{\sigma}$ are disordered in the gapped phase~\citep{jin2022unveiling}. Indeed, PMFT estimates a flux gap $\Delta_{\text{flux}}\apprle10^{-6}$ for $J_{z}\gtrsim1.2$, an energy difference that is smaller than the truncation error of DMRG simulations with 4000 kept states.

The $S=3/2$ KSL for $0<J_{z}<1$ and $D_{z}=0$ is gapless, characterized by a negative $Q^{z}$, and can be directly related to the $S=1/2$ gapless KSL. Recall the discussion in Sec.~\ref{sec:Conserved-Quantities}, where we showed how the $S=3/2$ KHM is projected onto the $S=1/2$ KHM with renormalized $J_{z}$ when $D_{z}\rightarrow +\infty$. The gapless $S=1/2$ KSLs is then adiabatically connected, e.g. without opening a gap,  to the $S=3/2$ KSL phases along the path in the $\left(J_{z},D_{z}\right)$ region.

In the $D_{z}\rightarrow\infty$ limit, the point $J_{z}=8$ marks the phase transition between the gapless and the gapped $S=1/2$ KHM phases, as shown in Fig.~\ref{fig:phasediagramNC}. This $S=1/2$ gapped phase is not adiabatically connected to the $S=3/2$ discussed above, since they are characterized by $Q^{z}$ parameters with different signs and any path connecting these phases in the $\left(J_{z},D_{z}\right)$ parameter space passes through a first-order quantum phase transition. 

\section{Effect of Out-of-Plane Single-ion anisotropy \label{sec:CSL}}
In this section, we study the $S=3/2$ KHM perturbed by an experimentally relevant out-of-plane SIA. We find that the resulting QSL breaks TRS and displays topologically nontrivial bands which are reminiscent of the chiral QSL of the KHM where it is induced by an out-of-plane magnetic field applied to the gapless $S=1/2$ KSL~\citep{Kitaev2006}. In the present case, TRS-breaking occurs spontaneously without an external magnetic field similar to cases of SU($N$) Heisenberg models in the large-$N$ limit~\citep{HermelePRB2011,YaoPRR2021,YaoPRB2022} or Kitaev models on graphs containing plaquettes with an odd number of vertices~\citep{YaoPRL2007,Chua2011,Natori2016,Natori2017,cassella2022exact}. 
However, we will show that in the case of the $S=3/2$ KHM, the sum of the Chern numbers is equal to zero, resulting in a non-chiral ground state.

\subsection{Three-spin interaction induced by single-ion anisotropy}

We now consider an out-of-plane SIA given by
\begin{equation}
H_{\text{SIA}}=-D_{c}\sum_{j}\left(S_{j}^{c}\right)^{2},\label{eq:111_SIA}
\end{equation}
in which the $c$ axis is indicated in Fig.~\ref{fig:mirrors}. Such a 
SIA is predicted to be relevant for the recently proposed $S=3/2$
Kitaev materials on the honeycomb lattice~\citep{XuNPJ2018,Xu2020,Stavropoulos2021}.
Moreover, Ref.~\citep{Xu2020} proposes that strain can tune the
van der Waals magnets into a model dominated by Kitaev interactions
and out-of-plane SIA. Therefore, Eq.~\eqref{eq:111_SIA}
is the simplest perturbation to the KHM, which has direct experimental implications. This term can be rewritten in terms of pseudo-dipoles
and pseudo-orbitals using Eq.~\eqref{eq:sigma_Ty} as follows
\begin{equation}
H_{\text{SIA}}=\frac{D_{c}}{3}\sum_{j}\left(\sigma_{j}^{x}+\sigma_{j}^{y}+\sigma_{j}^{z}\right)T_{j}^{y},\label{eq:H_SIA}
\end{equation}
in which we dropped off an unimportant constant. The presence of pseudo-dipoles in this expression shows that $H_{\text{SIA}}$ does not commute with $W_{p}^{\sigma}$ and creates flux excitations. Recent studies of the $S=1/2$ KHM have developed a piece of machinery to study non-flux-conserving
perturbations using variational methods~\citep{ZhangVariational2021,Zhang111Field2021}
or extensions of PMFT~\citep{Ralko2020,Knolle2018A,CookmeyerArxiv2022}. For simplicity, we will focus on the zero-flux sector within the third-order perturbation theory. 

The SIA induces a three spin-orbital interaction preserving the flux sector in analogy to the effect of a magnetic field on the $S=1/2$ KHM~\citep{Kitaev2006}. A straightforward way to show this is to rewrite Eq.~\eqref{eq:H_SIA} as
\begin{equation}
H_{\text{SIA}}=-\frac{D_{c}}{3}\sum_{j}i\left(\eta_{j}^{x}+\eta_{j}^{y}+\eta_{j}^{z}\right)\theta_{j}^{y},
\end{equation}
which is analogous to the representation of an applied magnetic field
on $S=1/2$ systems~\citep{Kitaev2006}. Notice that the only matter flavor involved in $H_{\text{SIA}}$ is $\theta^{y}$, indicating a direct influence on the flat bands. The third-order perturbation theory of $H_{\text{SIA}}$
displays a flux-conserving three-body interaction
\begin{align}
H^{(3)} & =\kappa\sum_{\left\langle ij\right\rangle _{\alpha}\left\langle jk\right\rangle _{\beta}}\left(\sigma_{i}^{\alpha}T_{i}^{y}\right)\left(\sigma_{j}^{\gamma}T_{j}^{y}\right)\left(\sigma_{k}^{\beta}T_{k}^{y}\right),\label{eq:H3_spin_orbital}
\end{align}
in which $i$ and $k$ are second-nearest neighbors, $j$ is the site bridging them, and $\kappa\sim\left(D_{c}/3\right)^{3}$. The SO(6) Majorana representation also provides an adequate representation of $H^{(3)}$, as it is clear by rewriting 
\begin{align}
\sigma_{i}^{\alpha}T_{i}^{y} & =-i\eta_{i}^{\alpha}\theta_{i}^{y},\nonumber \\
\sigma_{k}^{\beta}T_{k}^{y} & =-i\eta_{k}^{\beta}\theta_{k}^{y},\nonumber \\
\sigma_{j}^{\gamma}T_{j}^{y} & =\left(-i\eta_{j}^{\alpha}\eta_{j}^{\beta}\right)\left(-i\theta_{j}^{z}\theta_{j}^{x}\right),
\end{align}
which leads to 
\begin{align}
H^{(3)} & =-\kappa\sum_{\left\langle ij\right\rangle _{\alpha}\left\langle jk\right\rangle _{\beta}}\hat{U}_{\left\langle ik\right\rangle }\left(i\theta_{i}^{y}\theta_{j}^{z}\right)\left(i\theta_{j}^{x}\theta_{k}^{y}\right),\label{eq:H3_Majorana}
\end{align}
where where $\hat{U}_{\left\langle ik\right\rangle }=\hat{u}_{\left\langle ij\right\rangle _{\alpha}}\hat{u}_{\left\langle jk\right\rangle _{\beta}}$.

The general zero-flux mean-field decoupling of $H^{(3)}$ is given by
\begin{align}
H_{\text{MFT}}^{(3)} & =\kappa\sum_{\left\langle ij\right\rangle _{\alpha}\left\langle jk\right\rangle _{\beta}}\hat{U}_{\left\langle ik\right\rangle }\left[\Delta_{ij}^{yz}\left(i\theta_{j}^{x}\theta_{k}^{y}\right)+\Delta_{jk}^{xy}\left(i\theta_{i}^{y}\theta_{j}^{z}\right)\right]\nonumber \\
 & -\kappa\sum_{\left\langle ij\right\rangle _{\alpha}\left\langle jk\right\rangle _{\beta}}\hat{U}_{\left\langle ik\right\rangle }\left[\Delta_{ij}^{yx}\left(i\theta_{j}^{z}\theta_{k}^{y}\right)+\Delta_{jk}^{zy}\left(i\theta_{i}^{y}\theta_{j}^{x}\right)\right]\nonumber \\
 & +\kappa\sum_{\left\langle ij\right\rangle _{\alpha}\left\langle jk\right\rangle _{\beta}}\hat{U}_{\left\langle ik\right\rangle }\left[\xi_{ik}^{yy}\left(i\theta_{j}^{z}\theta_{j}^{x}\right)+Q_{j}^{y}\left(i\theta_{i}^{y}\theta_{k}^{y}\right)\right],\label{eq:H3_MFT}
\end{align}
in which we introduced second-nearest neighbor order parameters
\begin{align}
\xi_{ik}^{yy} & =-\left\langle i\theta_{i}^{y}\theta_{k}^{y}\right\rangle.
\end{align}

\begin{figure}
\begin{centering}
\includegraphics[width=0.5\columnwidth]{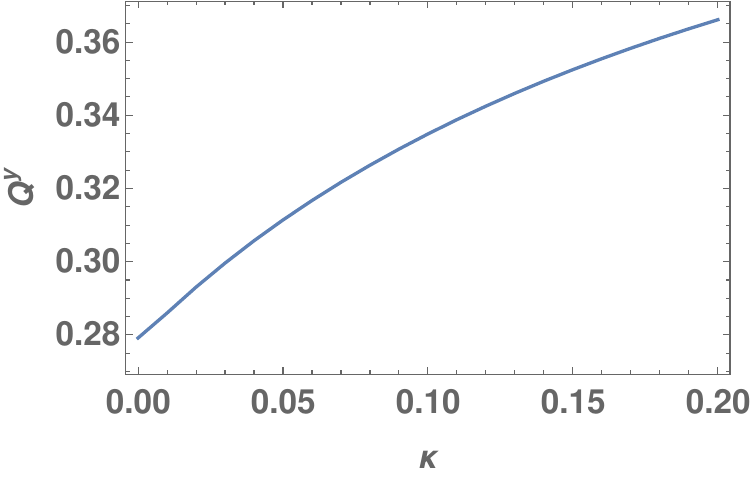}
\par\end{centering}
\caption{\label{fig:Qy_against_kappa} Evolution of the octupolar parameter $Q^{y}$ as a function of the three-site interaction quantified by $\kappa$ for $\kappa\in\left(0,0.2\right]$. The $\kappa = 0$ point marks a strong first-order phase transition that is followed by a smooth increase of $Q^{y}$. }
\end{figure}

A nonzero $\kappa$ in Eq.~\eqref{eq:H3_MFT} provides a positive feedback loop involving the formation of an octupolar order parameter $Q^{y}$ and the onset of second-nearest neighbor hoppings between $\theta_{i}^{y}$ particles. This implies that the isotropic $S=3/2$ KSL is unstable to breaking time-reversal symmetry under the influence of $H^{(3)}$. Since $Q^{y}\neq0$ implies time-reversal symmetry breaking, parameters such as $\Delta_{ij}^{yx}$ and $\Delta_{ij}^{yz}$ can now acquire nonzero values and enhance the hybridization between
$\theta^{y}$ flat band states and itinerant Majorana fermions. The complete hybridization of the low-energy flat bands leads to the first-order phase transition indicated in Fig.~\ref{fig:Qy_against_kappa}. For $\kappa=0.001$, we find that $Q_{A}^{y}=Q_{B}^{y}\approx0.28$ and second nearest-neighbor hopping parameters $\xi_{\mathbf{r},\mathbf{r}+\mathbf{d}_{\alpha},A}^{yy}=-\xi_{\mathbf{r},\mathbf{r}+\mathbf{d}_{\alpha},B}^{yy}\approx-0.115$, in which $\mathbf{d}_{\alpha=1,3,5}$ is indicated in Fig.~\ref{fig:plaquette}. A small value of $\kappa$ also leads to a large difference between the dispersion of the isotropic model in Fig.~\ref{fig:Representative_cases_3_2KSL}(a)
and the CSL dispersion in Fig.~\ref{fig:bands_S32_CSL}(a).

\subsection{Topological Properties of the Time-Reversal Symmetry Breaking Spin Liquid}

\begin{figure}
\begin{centering}
\includegraphics[width=0.5\columnwidth]{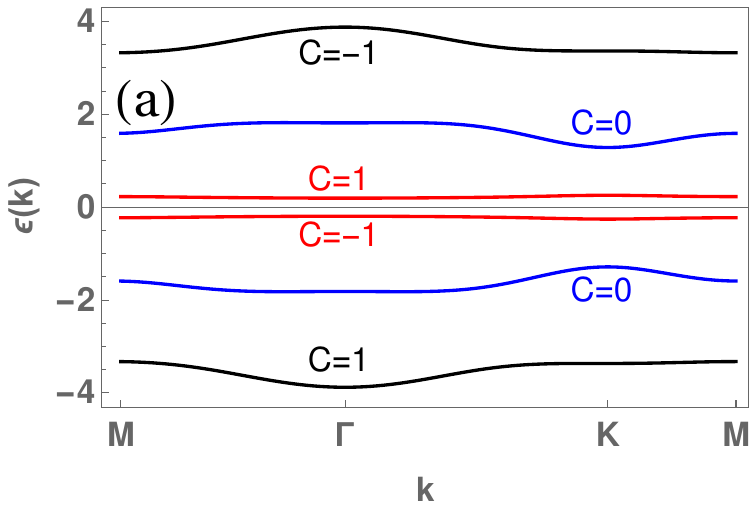}\includegraphics[width=0.5\columnwidth]{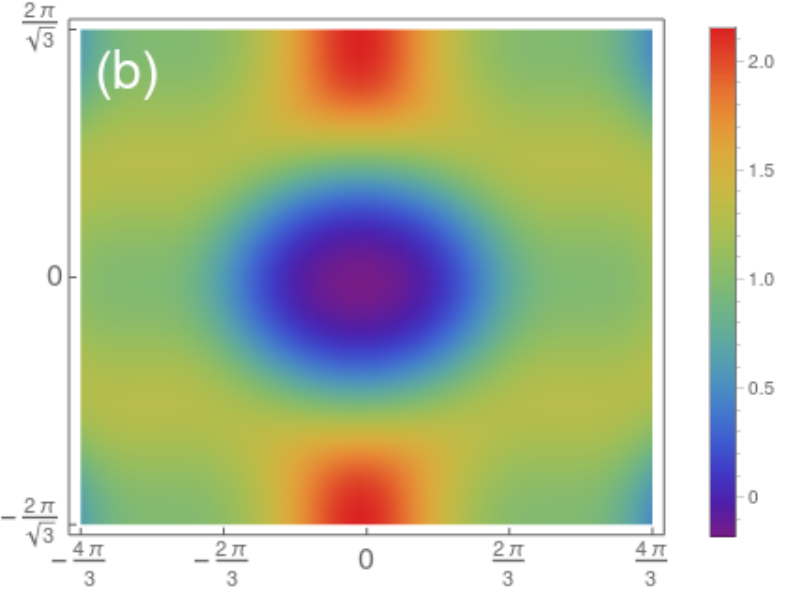}
\par\end{centering}
\begin{centering}
\includegraphics[width=0.5\columnwidth]{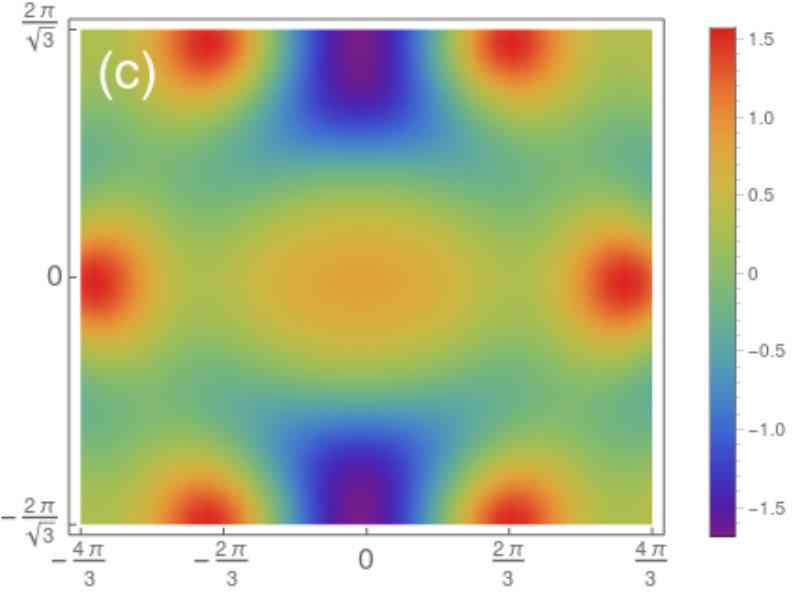}\includegraphics[width=0.5\columnwidth]{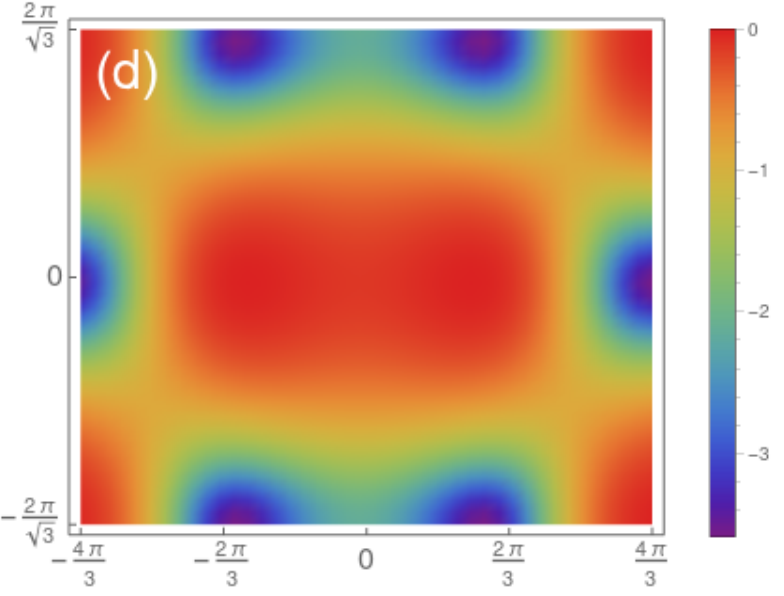}
\par\end{centering}
\caption{\label{fig:bands_S32_CSL} (a) Band structure of the spontaneously TRS breaking QSL at $\kappa=0.001$. Notice that all bands display narrow dispersions. Panels (b-d) display the z-component of the Berry curvature $\boldsymbol{\Omega}$, in which (b) corresponds to the lowest-energy band, (c) to the intermediate-energy band, and (d) to the highest-energy band. }
\end{figure}

Next, we discuss the topological properties of the TRS breaking $S=3/2$ KSL. After the sudden jump of the octupolar order parameter for infinitesimal $\kappa$ it grows slowly; for concreteness,  we fix $\kappa=0.001$. In this case, the CSL is characterized by three narrow bands, in which
the one closer to zero is particularly flat, see Fig.~\ref{fig:bands_S32_CSL}(a).

Their topological properties can be quantified by the Berry curvature
\begin{align}
\boldsymbol{\Omega}_{n}\left(\mathbf{k}\right) & =\nabla_{\mathbf{k}}\times\mathbf{A}_{n}\left(\mathbf{k}\right), 
\end{align}
in which $\mathbf{A}_{n}\left(\mathbf{k}\right)=i\left\langle u_{n}\left(\mathbf{k}\right)\left|\nabla_{\mathbf{k}}\right|u_{n}\left(\mathbf{k}\right)\right\rangle $
is the Berry connection of the $n$-th eigenstate $\left|u_{n}\left(\mathbf{k}\right)\right\rangle $
labeled by the wavevector $\mathbf{k}$. We computed the Berry curvature~\citep{Fukui2005}
and Figs.~\ref{fig:bands_S32_CSL}(b)-(d) displays the density plot
of the $z$ direction of $\boldsymbol{\Omega}_{n}\left(\mathbf{k}\right)$
of the negative energy bands. We compute the Chern number of the three negative energy bands
\begin{equation}
C_{n}=\frac{1}{2\pi}\int_{\text{BZ}}d^{2}\mathbf{k}\Omega_{n}^{z}\left(\mathbf{k}\right),
\end{equation}
and checked that bands with opposite energy dispersion display opposite Chern numbers. The lowest, intermediate, and highest energy bands have Chern numbers $C=1$, $C=0$, and $C=-1$, respectively. Hence, two of the bands are topologically nontrivial but the whole system has a total Chern number equal to zero. Therefore, no chiral edge mode crosses the gap around zero energy and the system is not a CSL. 

\begin{figure*}
\centering{}\includegraphics[width=0.5\columnwidth]{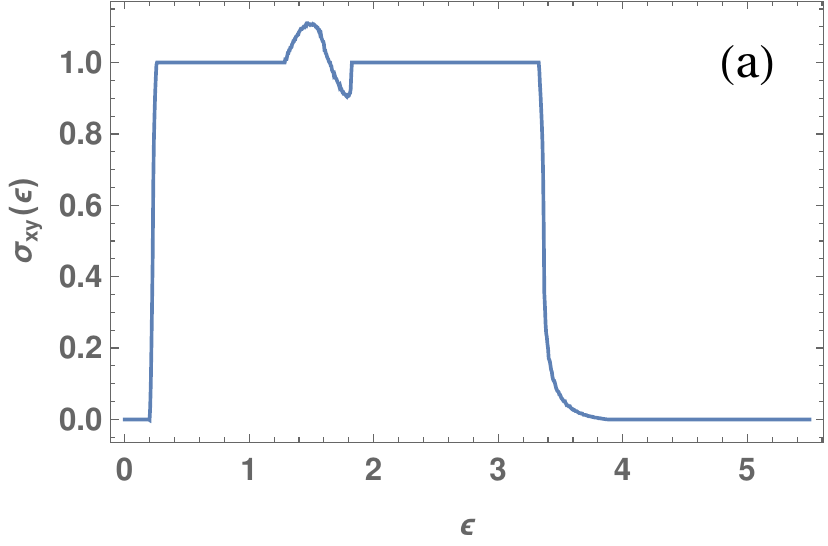}\includegraphics[width=0.5\columnwidth]{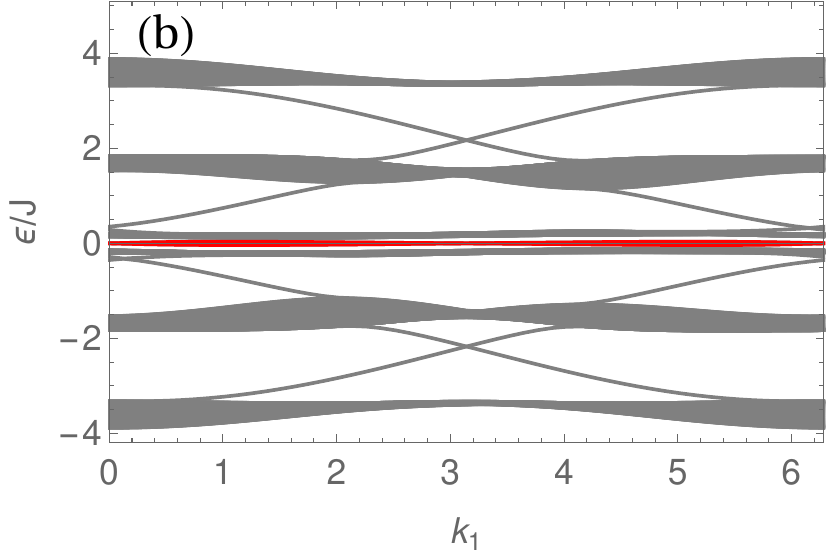}\includegraphics[width=0.5\columnwidth]{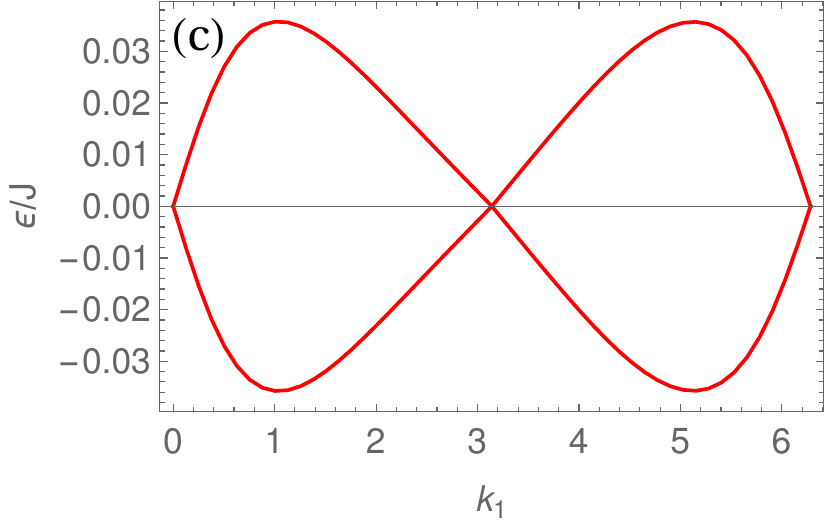}\includegraphics[width=0.5\columnwidth]{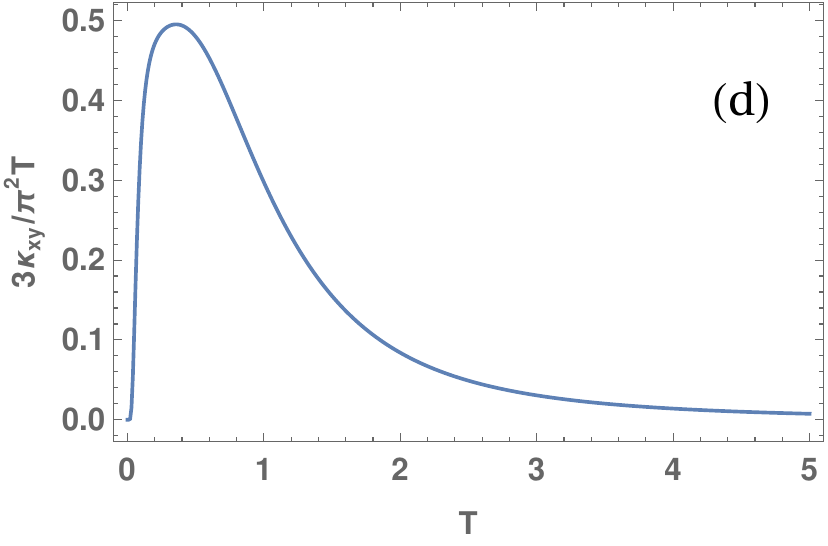}\caption{\label{fig:S32_CSL_edge_states} Topological characterization of the TRS breaking QSL. Panel (a) displays its Hall conductivity $\sigma(\epsilon)$ evaluated according to Eq. (\ref{eq:Hall_cond}). Panel (b) shows the QSL dispersion in open boundary conditions highlighting emergent low-energy edge states in red that are detailed in Panel (c). Panel (d) displays the thermal Hall conductivity $\kappa_{H}(T)/T $ and indicates a peak tending to the characteristic half-quantization value before decreasing monotonically with increasing temperature. }
\end{figure*}

Another function that also illustrates the non-trivial properties of the band topology is the Hall conductivity
$\sigma(\epsilon)$~\citep{Zhang_NPJ2016}
\begin{equation}
\sigma(\epsilon)=\frac{1}{V}\sum_{\mathbf{k},\epsilon_{\mathbf{k}}<\epsilon}\Omega^{z}(\mathbf{k}),\label{eq:Hall_cond}
\end{equation}
which is indicated in Fig.~\ref{fig:S32_CSL_edge_states}(a). Since the Majorana bands are gapped, $\sigma(\epsilon)=0$ for low energies. Then it jumps to $\sigma(\epsilon)=1$ due to the integration of $\Omega^{z}(\mathbf{k})$ of the lowest positive-energy band, as expected from its Chern number indicated in Fig.~\ref{fig:bands_S32_CSL}.
The Hall conductivity is kept constant in the gap between the lowest and second-lowest positive-energy bands. After reaching the second band, $\sigma(\epsilon)$ oscillates in accordance to the nonzero values of $\Omega^{z}(\mathbf{k})$, then returning to $\sigma(\epsilon)=1$. Finally, $\sigma(\epsilon)$ drops sharply to zero as the integration
occurs at the highest energy band. The non-trivial topological features in periodic boundary conditions are reflected by the existence of edge states in open boundary conditions, as indicated in Fig.~\ref{fig:S32_CSL_edge_states}(b). In this case, high-energy modes connect the two topologically nontrivial bands. Low-energy edge modes are also observed in Fig.~\ref{fig:S32_CSL_edge_states}(c) but they do not connect the bands and are topologically trivial.  

The standard signature for edge states in CSLs is the thermal Hall conductivity, which displays half-quantization due to the presence of zero-energy chiral Majorana edge states~\citep{Kitaev2006,Kasahara2018,Yokoi2021}.
For a flux-fixed background, we can estimate the thermal Hall conductivity through~\citep{Zhang_NPJ2016}
\begin{equation}
\kappa_{H}(T)=-\frac{1}{T}\int_{0}^{\infty}d\epsilon\,\epsilon^{2}\sigma(\epsilon)\frac{\partial f}{\partial\epsilon}(\epsilon,T),
\end{equation}
in which $f\left(\epsilon,T\right)$ is the Fermi-Dirac distribution.
Fig.~\ref{fig:S32_CSL_edge_states}(d) shows the numerically computed $\kappa_{H}(T)/T$. In contrast to CSLs, it vanishes at low temperatures and then  rapidly grows to a peak at a temperature scale when the chiral edge modes between the higher energy bands are thermally populated, which is similar to the behavior of topological magnon insulators. The value of the peak can still be quantified in terms of the thermal Hall conductivity of the chiral $S=1/2$ KSL, which reads~\citep{Kasahara2018,Yokoi2021}
\begin{equation}
\frac{\kappa_{\text{KSL}}^{1/2}}{T}=\frac{1}{2}\left(\frac{\pi^{2}k_{B}^{2}}{3\hbar}\right)C_{h},
\end{equation}
in which $C_{h}=\pm1$ according to the direction of the applied magnetic field. In contrast to the chiral $S=1/2$ KSL, the TRS breaking QSL discussed here does not reach the plateau, as indicated in Fig.~\ref{fig:S32_CSL_edge_states}(d).

\section{Conclusions and Outlook \label{sec:Conclusions}}

In this work, we have provided a detailed study of the $S=3/2$ KHM emphasizing its similarities with and relations to exactly solvable KK models~\citep{Yao2009,Nussinov2009,Yao2011,Chua2011,Farias2020,Natori2020,Chulliparambil2020,Seifert2020,Ray2021,Chulliparambil2021,Zhuang2021,WangHaoranPRB2021,Verresen2022}.
Our analysis mapped out the local symmetries of the model and analyzed the nature of the $S=3/2$ KSL phases. We showed that the model still contains an exact static Z$_2$ gauge field and in a given flux sector it is a sum of bilinear Majorana operators and quartic and sextic interactions. The presence of an exactly soluble part of the $S=3/2$ KHM, e.g. a kinetic term before the parton mean-field decoupling, also rationalizes the remarkable quantitative agreement between PMFT and DMRG simulations found previously~\citep{jin2022unveiling}. 

The symmetry analysis was crucial for understanding the first-order phase transition occurring when introducing anisotropies in the couplings. Namely,  it provides tight constraints for the order parameters and shows the emergence of a low-energy Majorana  flat band. The pseudo-dipole
and pseudo-orbital operators in a KK-like representation of the model were useful for uncovering similarities between the $S=3/2$ {[}111{]} SIA and the $S=1/2$ out-of-plane magnetic field. The latter motivated us to study the $S=3/2$ KHM with this experimentally relevant SIA and we argue that the system displays spontaneous TRS breaking. Some of the Majorana bands of resulting QSL acquire nonzero Chern numbers but the TRS phase is different from the standard chiral QSL because the sum over the Chern number of all bands is zero. Hence, no quantization of the thermal Hall conductivity is expected at very low temperatures but only a broad maximum at finite temperatures. 

Our work opens a number of avenues for future research. It would be interesting to verify if the techniques that we apply for the $S=3/2$ KHM in this paper can be generalized for higher-spin systems with $S=(2^{n}-1)/2$ ($n\in\mathbb{N}$), as suggested by the exactly solvable models discussed in Section~\ref{sec:Conserved-Quantities}. We foresee that such a study can provide a complementary approach to the large-$S$ limit of this model~\citep{IoannisNatComm2018} but within a natural extension of Kitaev's original formalism~\citep{Kitaev2006}. It would also be consistent with a recent study showing that half-integer KHMs always display deconfined $Z_2$ fermionic gauge charges \citep{MaPRL2023}. Another open problem concerns the systematic study of the $S=3/2$ KHM in different flux sectors, in the presence of disorder or vacancies. The introduction of flux excitations would also allow the computation of different dynamical response functions for experimental detection following Ref.~\citep{Natori2020}.  

Finally, it would be very worthwhile to systematically study implementations of the $S=3/2$ KHM in van der Waals magnets. Studies using ab initio\textcolor{red}{{} }\textcolor{black}~{\citep{XuNPJ2018,Xu2020}
and quantum chemistry~\citep{Stavropoulos2021} methods suggest that the Kitaev exchange is present in van der Waals ferromagnets such as CrI$_{3}$ and CrXTe$_{3}$ (X=Si,Ge) due to their ligands strong spin-orbit coupling. The theoretical studies indicate that the Kitaev interaction should be substantially smaller than the Heisenberg one, a result that is consistent with the data from a recent neutron scattering
experiment on CrI$_{3}$~\citep{Chen2021}. However, the same theories also suggest that strain can dramatically change the exchange constants, and even induce a model dominated by Kitaev interactions and {[}111{]} SIA~\citep{Xu2020}. This strain is experimentally feasible, as it can be applied mechanically or by proximity effects in metal-insulator
heterostructures}\textcolor{red}{{} }\textcolor{black}~{\citep{Biswas2019,Leeb2021}.
When combined with better strategies for quantifying exchange constants~\citep{Cen2022}, microscopic studies can help to discover new QSL candidates in higher-spin and spin-orbital systems.}

\section*{Acknowledgments}
We thank F. Pollmann for important discussions and collaboration on previous related work. 
W.N. would like to thank F. Alcaraz for suggesting a connection to parafermions, and to R. Pereira, E. Andrade, and E. Miranda for works on related projects. W.N. also thanks T. Ziman and M. Zhitomirsky for discussions
about van der Waals magnets. 

W.N. and J.K.  acknowledge the support of the Royal Society via a Newton International Fellowship through project NIF-R1-181696, during which many of the results in the manuscript were derived. H.-K. J. is funded by the European Research Council (ERC) under the European Unions Horizon 2020 research and innovation program (grant agreement No. 771537).

JK is part of the Munich Quantum Valley, which is supported by the Bavarian state government with funds from the Hightech Agenda Bayern Plus.
We also acknowledge the support of the Imperial-TUM flagship partnership.

\bibliographystyle{apsrev4-2}
\addcontentsline{toc}{section}{\refname}\bibliography{j32_KM}

\appendix

\section{Pseudo-dipolar and Pseudo-orbital Matrix Representation on the basis
of $S=3/2$ spins\label{sec:Matrices_sigma_T}}

The equivalence between $\left|S=\frac{3}{2},S^{z}\right\rangle $
vectors and vectors $\left|s^{z}=\frac{1}{2}\sigma^{z},\tau^{z}=\frac{1}{2}T^{z}\right\rangle $
can be written like
\begin{align}
\left|S=\frac{3}{2},S^{z}=\frac{3}{2}\right\rangle  & =\left|s^{z}=-\frac{1}{2},\tau^{z}=\frac{1}{2}\right\rangle ,\nonumber \\
\left|S=\frac{3}{2},S^{z}=\frac{1}{2}\right\rangle  & =-\left|s^{z}=\frac{1}{2},\tau^{z}=-\frac{1}{2}\right\rangle ,\nonumber \\
\left|S=\frac{3}{2},S^{z}=-\frac{1}{2}\right\rangle  & =\left|s^{z}=-\frac{1}{2},\tau^{z}=-\frac{1}{2}\right\rangle ,\nonumber \\
\left|S=\frac{3}{2},S^{z}=-\frac{3}{2}\right\rangle  & =-\left|s^{z}=\frac{1}{2},\tau^{z}=\frac{1}{2}\right\rangle .
\end{align}
In the ordered basis $\left\{ \left|S=\frac{3}{2},S^{z}\right\rangle \right\} $
with decreasing $S^{z}$, the $\sigma^{\gamma}$ operators are represented
by
\begin{align}
\boldsymbol{\sigma} & =\left(-\rho^{x}\otimes\rho^{x},\rho^{x}\otimes\rho^{y},-\rho^{0}\otimes\rho^{z}\right),\nonumber \\
\mathbf{T} & =\left(\rho^{x}\otimes\rho^{0},\rho^{y}\otimes\rho^{z},\rho^{z}\otimes\rho^{z}\right),
\end{align}
in which $\rho^{\gamma}$ are the Pauli matrices and $\rho^{0}$ is
the $2\times2$ identity. 

\section{Jordan-Wigner Transformation of the $S=3/2$ Kitaev Model \label{sec:JWT_S32}}

In this appendix, we discuss the JWT introduced in Ref.~\citep{Baskaran2008}
using the pseudo-dipole and pseudo-orbitals operators. Let us label
the sites on the honeycomb lattice according to their positions along
the $xy$ chains with the indexes $(l,m)$, in which $l+m$ even (odd)
corresponds to points on the $A$ ($B$) sublattice, see Fig.~\ref{fig:plaquette}(b).
The JWT represents the spin operator by combining a string of spins
with a fermionic operator $c_{(l,m)}^{(\dagger)}$ on the edge as
follows
\begin{align}
\sigma_{(l,m)}^{+} & =\left[\prod_{m^{\prime}<m}\prod_{l^{\prime}}\sigma_{\left(l^{\prime},m^{\prime}\right)}^{z}\right]\left[\prod_{l^{\prime}<l}\sigma_{\left(l^{\prime},m\right)}^{z}\right]c_{(l,m)}^{\dagger},\nonumber \\
\sigma_{(l,m)}^{z} & =2c_{(l,m)}^{\dagger}c_{(l,m)}-1.\label{eq:JWT_sigma}
\end{align}
The $c_{(l,m)}$ (canonical) fermions are conveniently combined into
the Majorana fermions
\begin{align}
\theta_{(l,m)}^{0}= & \begin{cases}
\frac{1}{i}\left(c_{(l,m)}-c_{(l,m)}^{\dagger}\right), & \text{ if }l+m\text{ is even},\\
c_{(l,m)}+c_{(l,m)}^{\dagger}, & \text{ if }l+m\text{ is odd}.
\end{cases}\nonumber \\
\chi_{(l,m)}= & \begin{cases}
c_{(l,m)}+c_{(l,m)}^{\dagger}, & \text{ if }l+m\text{ is even}.\\
\frac{1}{i}\left(c_{(l,m)}-c_{(l,m)}^{\dagger}\right), & \text{ if }l+m\text{ is odd},
\end{cases}\label{eq:Majorana_sigma}
\end{align}
which are used to represent the matter and gauge sectors, respectively.
It is straightforward to prove that they satisfy the algebra
\begin{align}
\left\{ \theta_{(l,m)}^{0},\theta_{(r,s)}^{0}\right\} =\left\{ \chi_{(l,m)},\chi_{(r,s)}\right\}  & =2\delta_{(l,m),(r,s)},\nonumber \\
\left\{ \theta_{(l,m)}^{0},\chi_{(r,s)}^{0}\right\}  & =0.\label{eq:algebra_theta0}
\end{align}
For simplicity, we set the following notation for the nearest neighbors
\begin{subequations}
\begin{align}
\left(r,s\right)_{\gamma} & =\begin{cases}
\left(r+1,s\right), & \text{if }\gamma=x,\\
\left(r-1,s\right), & \text{if }\gamma=y,\\
\left(r,s-1\right), & \text{if }\gamma=z,
\end{cases}
\end{align}
as well as a bond operator
\begin{equation}
\hat{\mu}_{\left(r,s\right)}^{\gamma}=\begin{cases}
-i\chi_{(r,s)}\chi_{(r,s-1)}, & \text{if }\gamma=z,\\
1, & \text{otherwise}.
\end{cases}
\end{equation}
\end{subequations} In terms of the equations above, the model $H_{\text{Kit}}^{\sigma}$
defined on the main text reads
\begin{align}
H_{\text{Kit}}^{\sigma} & =\frac{i}{4}\sum_{l+m\text{ even}}J_{\gamma}\hat{\mu}_{\left(l,m\right)}^{\gamma}\theta_{\left(l,m\right)}^{0}\theta_{\left(l,m\right)_{\gamma}}^{0}.\label{eq:H-Kit-1_2}
\end{align}
Since$\hat{\mu}_{\left(l,m\right)}^{2}=1$ and $\left[H_{\text{Kit}}^{\sigma},\hat{\mu}_{\left(l,m\right)}\right]=0$,
$\hat{\mu}_{\left(r,s\right)}^{z}$ can be regarded as a $Z_{2}$ bond operator that can be fixed. Such operation is equivalent to fixing
the eigenstates $W_{p}^{\sigma}$ in Eq.~\eqref{eq:Wp_sigma}, thus
defining the flux sector. Notice that the SO(6) Majorana partons lead
to the same mapping, but with extra gauge variables on the $x$ and $y$ bonds. 

We need to include the pseudo-dipoles in order to complement the JWT
defined in Eq.~\eqref{eq:JWT_sigma}. For this purpose, it is convenient
to represent $T^{\gamma}$ using hard-core bosons $\left(d_{(l,m)},d_{(l,m)}^{\dagger}\right)$
at each site of the lattice as follows \citep{MatsubaraMatsuda1956,Wu2002}
\begin{align}
\left[d_{(l,m)},d_{(r,s)}\right]=\left[d_{(l,m)},d_{(r,s)}^{\dagger}\right] & =0,\text{ if }(l,m)\neq(r,s),\nonumber \\
\left\{ d_{(r,s)},d_{(r,s)}^{\dagger}\right\}  & =1,\nonumber \\
\left\{ d_{(r,s)},d_{(r,s)}\right\} =\left\{ d_{(r,s)}^{\dagger},d_{(r,s)}^{\dagger}\right\}  & =0.
\end{align}
The isomorphism between the hard-core boson Fock space and the pseudo-orbital
operators are ensured by the relations 
\begin{align}
T_{(l,m)}^{y} & =1-2d_{(l,m)}^{\dagger}d_{(l,m)},\nonumber \\
T_{(l,m)}^{z} & =d_{(l,m)}^{\dagger}+d_{(l,m)},\nonumber \\
T_{(l,m)}^{x} & =i\left(d_{(l,m)}^{\dagger}-d_{(l,m)}\right),\label{eq:T_hard-core_bosons}
\end{align}
in which we settled an equivalence between $\left|n_{d}\right\rangle $
and eigenstates of $T^{y}$. Since $\left[\sigma_{(r,s)}^{\alpha},T_{(l,m)}^{\beta}\right]=0$,
we also demand that the hard-core bosons commute with the Majorana
fermions in Eq.~\eqref{eq:algebra_theta0}. By mapping the orbitals
into a Fock space through Eq.~\eqref{eq:T_hard-core_bosons}, one
can define the operators $\theta_{(l,m)}^{\alpha}$
\begin{align}
\theta_{(l,m)}^{\gamma} & =\theta_{(l,m)}^{0}T_{(l,m)}^{\gamma}.\label{eq:parafermion_version_theta_gamma}
\end{align}
Notice that the Hilbert space of $\theta^{\gamma}$ is two times larger
than the one of $\theta^{0}$ and that it is shared by the three operators
defined in Eq.~\eqref{eq:parafermion_version_theta_gamma}. In a model
that retains both $\theta^{0}$ and $\theta^{\alpha}$, one should
bear in mind that the identity $T^{x}T^{y}T^{z}=i$ leads to the constraint
\begin{equation}
\theta_{(l,m)}^{0}=-i\theta_{(l,m)}^{x}\theta_{(l,m)}^{y}\theta_{(l,m)}^{z},\label{eq:theta_constraint}
\end{equation}
which is exactly the same as we derived in terms of SO(6) partons. 

The algebraic relations of the operators derived in this appendix
satisfy all properties expected for Majorana fermions, except for
the same-site \emph{commutation} relation between $\theta_{i}^{0}$
and $\theta_{i}^{\alpha}$. More explicitly,
\begin{align}
\left\{ \theta_{(l,m)}^{\alpha},\theta_{(r,s)}^{\beta}\right\}  & =2\delta^{\alpha\beta}\delta_{(l,m),(r,s)},\nonumber \\
\left\{ \theta_{(l,m)}^{0},\theta_{(r,s)}^{\alpha}\right\}  & =0,\text{if }(l,m)\neq(r,s)\nonumber \\
\left[\theta_{i}^{0},\theta_{i}^{\alpha}\right] & =0.\label{eq:theta_JWT}
\end{align}
The mixture of bosonic and fermionic properties in Eq.~\eqref{eq:theta_JWT}
is reminiscent of the concept of parastatistics introduced by Green~\citep{Green1953}, whose original interest was to generalize the
method of second quantization and demonstrate the theoretical possibility
of free particles that do not obey the usual symmetrization principles.
Green's parafermions are characterized by a field $a_{k}$ that is divided into $p$ components~\citep{Green1953,GreenbergMessiah1965}
that is represented like~\citep{Macfarlane1994}: 
\begin{equation}
a_{k}=\sum_{\alpha=1}^{p}f_{k}^{(\alpha)}\xi_{k}^{(\alpha)},\label{eq:GreenParafermions}
\end{equation}
in which $f_{k}^{(\alpha)}$ is a canonical fermion (boson) for parafermions
(parabosons), and $\xi_{k}^{(\alpha)}$ is a Majorana fermion. We
can then see that $\theta^{\alpha}$ is not a Green's parafermion
since it is constructed by a combination of a Majorana fermion and
operators in terms of hard-core bosons. Given this qualification,
$\theta^{0}$ and $\theta^{\alpha}$ follow the spirit of Green's original parafermions by displaying algebraic properties that are neither bosonic nor fermionic. Another usage of the term parafermion
refers to the $Z_{k}$-clock generalizations of Majorana fermions~\citep{Vaezi2014,Barkeshli2015,Alicea2016,Fendley2014}. Although
$\theta^{\gamma}$ is not within this class of operators, they can
be also understood as a Majorana fermion generalization. 

\section{Projection operator of SO(6) Majorana fermions \label{sec:Projection}}

Let us discuss the explicit formula for the $D$ operator in Eq. (\ref{eq:projector2}). It is convenient to define the matter fermions in terms of fermionic operators with well-defined occupation numbers such as 
\begin{align}
f_{\mathbf{r}}^{\gamma} & =\frac{\eta_{\mathbf{r}A}^{\gamma}+i\eta_{\mathbf{r}B}^{\gamma}}{2}.
\end{align}
A closed formula for the projector $P$ can be exactly derived in this case by showing that $D$ is given by~\citep{Pedrocchi2011,Zschocke2015}
\begin{equation}
D=\left(-1\right)^{\theta}\det Q^{u}\hat{\pi}\prod_{\left\langle ij\right\rangle _{\gamma}}u_{\left\langle ij\right\rangle _{\gamma}},
\end{equation}
in which $\theta$ is the function of the lattice boundary conditions derived in Refs.~\citep{Pedrocchi2011,Zschocke2015}, and $\hat{\pi}$
is the parity of the $f_{\mathbf{r}}^{\gamma}$ occupation numbers. The matrix $Q^{u}$ relates the Majorana fermions $\left\{ \theta_{i}^{\gamma}\right\} $
and the matter eigenstates at a fixed flux, and the product $\det Q^{u}\prod_{\left\langle ij\right\rangle _{\gamma}}u_{\left\langle ij\right\rangle _{\gamma}}$
is gauge invariant. The projection operator selects $\left|\psi_{0}\right\rangle $ states satisfying a parity condition of bond and then performing an equal weight linear superposition of all gauge transformations acting
on$\left|\psi_{0}\right\rangle $~\citep{YaoPRL2007}.

\section{Mean-Field Decoupling of the $S=3/2$ KHM \label{sec:MFT-Decoupling}}

Following Eqs.~\eqref{eq:S_gamma}, \eqref{eq:Majorana_parton}, \eqref{eq:theta_0_parton},
and \eqref{eq:theta_0}, the $S=3/2$ spin operators are represented
by~\citep{jin2022unveiling}
\begin{equation}
S^{\gamma}=\frac{i}{2}\eta^{\gamma}\left(\theta^{0}+2\theta^{\alpha\beta}\right).\label{eq:S_parton}
\end{equation}
Let us write the sites of the honeycomb lattice using a two-site basis
on the triangular lattice, for which the KHM reads 
\begin{align}
H_{\text{Kit}} & =\sum_{\mathbf{r}}\frac{iJ_{\gamma}}{4}\hat{u}_{\mathbf{r}A;\mathbf{r}_{\gamma}B}\left(\theta_{\mathbf{r}A}^{0}+2\theta_{\mathbf{r}A}^{\alpha\beta}\right)\left(\theta_{\mathbf{r}_{\gamma}B}^{0}+2\theta_{\mathbf{r}_{\gamma}B}^{\alpha\beta}\right),\label{eq:parton_S32_KHM}
\end{align}
in which $\mathbf{r}_{\gamma}=\mathbf{r}+\mathbf{a}_{\gamma}$, with
$\mathbf{a}_{z}=\mathbf{0}$. The exactly solvable model $H_{\text{Kit}}^{\sigma T}$
requires no mean-field decoupling and its mapping to a free fermion
problem is still given by Eq.~\eqref{eq:HsigTparton}. The non-integrable
model $H_{\text{Kit}}^{\sigma,\sigma T}$ is quartic in terms of SO(6)
partons and its most general decoupling given by
\begin{align}
H_{\text{Kit},\text{MFT}}^{\sigma,\sigma T}= & \sum_{\mathbf{r},\gamma,\lambda}\frac{J_{\gamma}}{2}u_{\mathbf{r}A;\mathbf{r}_{\gamma}B}\left(Q_{A}^{\lambda}i\theta_{\mathbf{r}A}^{\lambda}\theta_{\mathbf{r}_{\gamma}B}^{\alpha\beta}+Q_{B}^{\lambda}i\theta_{\mathbf{r}A}^{\alpha\beta}\theta_{\mathbf{r}_{\gamma}B}^{\lambda}\right)\nonumber \\
 & +\sum_{\mathbf{r},\gamma,\lambda}\frac{J_{\gamma}}{2}u_{\mathbf{r}A;\mathbf{r}_{\gamma}B}\Delta_{\mathbf{r}A,\mathbf{r}_{\gamma}B}^{\nu,\alpha\beta}i\theta_{\mathbf{r}A}^{\lambda}\theta_{\mathbf{r}A}^{\mu}\nonumber \\
 & +\sum_{\mathbf{r},\gamma,\lambda}\frac{J_{\gamma}}{2}u_{\mathbf{r}A;\mathbf{r}_{\gamma}B}\Delta_{\mathbf{r}A,\mathbf{r}_{\gamma}B}^{\alpha\beta,\nu}i\theta_{\mathbf{r}_{\gamma}B}^{\lambda}\theta_{\mathbf{r}_{\gamma}B}^{\mu},\label{eq:H4_MFT}
\end{align}
in which the Greek letters are specified by the anti-symmetric symbol
with $\epsilon^{\lambda\mu\nu}=1$. Finally, although $H_{\text{Kit}}^{\sigma}$
is integrable, the algebraic relation in Eq.~\eqref{eq:theta_0_comm}
requires that we perform a mean field decoupling. Its most general
decoupling is given by
\begin{align}
H_{\text{Kit},\text{MFT}}^{\sigma} & =\sum_{\mathbf{r},\gamma,\lambda,\lambda^{\prime}}t_{\mathbf{r}A;\mathbf{r}_{\gamma}B}^{(6),\lambda\lambda^{\prime}}u_{\mathbf{r}A;\mathbf{r}_{\gamma}B}i\theta_{\mathbf{r}A}^{\lambda}\theta_{\mathbf{r}_{\gamma}B}^{\lambda^{\prime}}\nonumber \\
 & +\sum_{\mathbf{r},\lambda}\sum_{X=A,B}t_{0,X}^{(6),\lambda\mu}\left(i\theta_{\mathbf{r}X}^{\lambda}\theta_{\mathbf{r}X}^{\mu}\right),\label{eq:H6_MFT}
\end{align}
in which 
\begin{align}
t_{\mathbf{r}A;\mathbf{r}_{\gamma}B}^{(6),\lambda\lambda^{\prime}} & =\frac{J_{\gamma}}{4}\left(Q_{\mathbf{r}A}^{\lambda}Q_{\mathbf{r}_{\gamma}B}^{\lambda^{\prime}}-\Delta_{\left\langle ij\right\rangle _{\gamma}}^{\mu\mu^{\prime}}\Delta_{\left\langle ij\right\rangle _{\gamma}}^{\nu\nu^{\prime}}+\Delta_{\left\langle ij\right\rangle _{\gamma}}^{\mu\nu^{\prime}}\Delta_{\left\langle ij\right\rangle _{\gamma}}^{\nu\mu^{\prime}}\right),\nonumber \\
t_{0,A}^{(6),\lambda\mu} & =\sum_{\gamma,\rho}\frac{J_{\gamma}}{4}\Delta_{\mathbf{r}A;\mathbf{r}_{\gamma}B}^{\nu\rho}Q_{\mathbf{r}_{\gamma}B}^{\rho}u_{\mathbf{r}A;\mathbf{r}_{\gamma}B},\nonumber \\
t_{0,B}^{(6),\lambda\mu} & =\sum_{\gamma,\rho}\frac{J_{\gamma}}{4}Q_{\mathbf{r}A}^{\rho}\Delta_{\mathbf{r}A;\mathbf{r}_{\gamma}B}^{\rho\nu}u_{\mathbf{r}A;\mathbf{r}_{\gamma}B}.\label{eq:PMFT_par_H_sig}
\end{align}
Once the flux sector is fixed, the order parameters in Eq.~\ref{eq:order_parameters}
are evaluated self-consistently. 

\section{$C_{3}$ symmetries and order parameters \label{sec:C3_relation_OP}}

In this appendix, we table the explicit relationships between the
order parameters $\Delta_{\left\langle ij\right\rangle _{\gamma}}^{ab}$
due to the $C_{3}$ rotation symmetry. For order parameters that do
not involve $\theta^{y}$ fermions, we find 
\begin{subequations}
\begin{equation}
\left(\begin{array}{c}
\Delta_{\left\langle ij\right\rangle _{x}}^{zz}\\
\Delta_{\left\langle ij\right\rangle _{x}}^{xx}\\
\Delta_{\left\langle ij\right\rangle _{x}}^{xz}\\
\Delta_{\left\langle ij\right\rangle _{x}}^{zx}
\end{array}\right)=\frac{1}{4}\left(\begin{array}{cccc}
1 & 3 & \sqrt{3} & \sqrt{3}\\
3 & 1 & -\sqrt{3} & -\sqrt{3}\\
-\sqrt{3} & \sqrt{3} & 1 & -3\\
-\sqrt{3} & \sqrt{3} & -3 & 1
\end{array}\right)\left(\begin{array}{c}
\Delta_{\left\langle ij\right\rangle _{z}}^{zz}\\
\Delta_{\left\langle ij\right\rangle _{z}}^{xx}\\
\Delta_{\left\langle ij\right\rangle _{z}}^{xz}\\
\Delta_{\left\langle ij\right\rangle _{z}}^{zx}
\end{array}\right),
\end{equation}
\begin{equation}
\left(\begin{array}{c}
\Delta_{\left\langle ij\right\rangle _{y}}^{zz}\\
\Delta_{\left\langle ij\right\rangle _{y}}^{xx}\\
\Delta_{\left\langle ij\right\rangle _{y}}^{xz}\\
\Delta_{\left\langle ij\right\rangle _{y}}^{zx}
\end{array}\right)=\frac{1}{4}\left(\begin{array}{cccc}
1 & 3 & -\sqrt{3} & -\sqrt{3}\\
3 & 1 & \sqrt{3} & \sqrt{3}\\
\sqrt{3} & -\sqrt{3} & 1 & -3\\
\sqrt{3} & -\sqrt{3} & -3 & 1
\end{array}\right)\left(\begin{array}{c}
\Delta_{\left\langle ij\right\rangle _{z}}^{zz}\\
\Delta_{\left\langle ij\right\rangle _{z}}^{xx}\\
\Delta_{\left\langle ij\right\rangle _{z}}^{xz}\\
\Delta_{\left\langle ij\right\rangle _{z}}^{zx}
\end{array}\right).
\end{equation}
\end{subequations} Conversely, if $\theta^{y}$ fermions are involved,
we find
\begin{subequations}
\begin{equation}
\left(\begin{array}{c}
\Delta_{\left\langle ij\right\rangle _{x}}^{xy}\\
\Delta_{\left\langle ij\right\rangle _{x}}^{zy}\\
\Delta_{\left\langle ij\right\rangle _{x}}^{yx}\\
\Delta_{\left\langle ij\right\rangle _{x}}^{yz}
\end{array}\right)=\frac{1}{2}\left(\begin{array}{cccc}
-1 & \sqrt{3}\\
-\sqrt{3} & -1\\
 &  & -1 & \sqrt{3}\\
 &  & -\sqrt{3} & -1
\end{array}\right)\left(\begin{array}{c}
\Delta_{\left\langle ij\right\rangle _{z}}^{xy}\\
\Delta_{\left\langle ij\right\rangle _{z}}^{zy}\\
\Delta_{\left\langle ij\right\rangle _{z}}^{yx}\\
\Delta_{\left\langle ij\right\rangle _{z}}^{yz}
\end{array}\right),
\end{equation}
\begin{equation}
\left(\begin{array}{c}
\Delta_{\left\langle ij\right\rangle _{y}}^{xy}\\
\Delta_{\left\langle ij\right\rangle _{y}}^{zy}\\
\Delta_{\left\langle ij\right\rangle _{y}}^{yx}\\
\Delta_{\left\langle ij\right\rangle _{y}}^{yz}
\end{array}\right)=\frac{1}{2}\left(\begin{array}{cccc}
-1 & -\sqrt{3}\\
\sqrt{3} & -1\\
 &  & -1 & -\sqrt{3}\\
 &  & \sqrt{3} & -1
\end{array}\right)\left(\begin{array}{c}
\Delta_{\left\langle ij\right\rangle _{z}}^{xy}\\
\Delta_{\left\langle ij\right\rangle _{z}}^{zy}\\
\Delta_{\left\langle ij\right\rangle _{z}}^{yx}\\
\Delta_{\left\langle ij\right\rangle _{z}}^{yz}
\end{array}\right).
\end{equation}
\end{subequations}

\end{document}